\documentclass[twocolumn]{aastex631}

\usepackage{graphicx}
\usepackage{amsmath}
\usepackage{natbib}
\usepackage{amssymb}

\newcommand{\be}{\begin{equation}}
\newcommand{\ee}{\end{equation}}

\newcommand{\appropto}{\mathrel{\vcenter{
		\offinterlineskip\halign{\hfil$##$\cr
	\propto\cr\noalign{\kern2pt}\sim\cr\noalign{\kern-2pt}}}}}

\shorttitle{The Effects of $r$-Process Enrichment in Hydrogen-Rich SNe}
\shortauthors{Patel, Goldberg, Renzo, Metzger}
\begin{document}

\title{The Effects of $r$-Process Enrichment in Hydrogen-Rich Supernovae}

\correspondingauthor{Anirudh Patel}
\email{anirudh.p@columbia.edu}

\author[0009-0000-1335-4412]{Anirudh Patel}
\affil{Department of Physics and Columbia Astrophysics Laboratory, Columbia University, New York, NY 10027, USA}

\author[0000-0003-1012-3031]{Jared A.~Goldberg}
\affil{Center for Computational Astrophysics, Flatiron Institute, 162 5th Ave, New York, NY 10010, USA}

\author[0000-0002-6718-9472]{Mathieu Renzo}
\affil{Steward Observatory, University of Arizona, 933 N. Cherry Avenue, Tucson, AZ 85721, USA}
\affil{Center for Computational Astrophysics, Flatiron Institute, 162 5th Ave, New York, NY 10010, USA}

\author[0000-0002-4670-7509]{Brian D.~Metzger}
\affil{Department of Physics and Columbia Astrophysics Laboratory, Columbia University, New York, NY 10027, USA}
\affil{Center for Computational Astrophysics, Flatiron Institute, 162 5th Ave, New York, NY 10010, USA}

\accepted{in ApJ, March 26, 2024}

\begin{abstract}
     Core-collapse supernovae are candidate sites for rapid neutron capture process ($r$-process) nucleosynthesis. We explore the effects of enrichment from $r$-process nuclei on the light-curves of hydrogen-rich supernovae (SNe IIP) and assess the detectability of these signatures. We modify the radiation hydrodynamics code $\texttt{SNEC}$ to include the approximate effects of opacity and radioactive heating from $r$-process elements in the SN ejecta. We present models spanning a range of total $r$-process masses $M_{\rm r}$ and their assumed radial distribution within the ejecta, finding that $M_{\rm r} \gtrsim 10^{-2} M_\odot$ is sufficient to induce appreciable differences in their light-curves as compared to ordinary SNe IIP (without any $r$-process elements). The primary photometric signatures of $r$-process enrichment include a shortening of the plateau phase, coinciding with the hydrogen-recombination photosphere retreating to the $r$-process-enriched layers, and a steeper post-plateau decline associated with a reddening of the SN colors.
     We compare our $r$-process-enriched models to ordinary IIP models and observational data, showing that yields of $M_{\rm r} \gtrsim 10^{-2} M_\odot$ are potentially detectable across several of the metrics used by transient observers, provided that $r$-process rich layers are mixed $\gtrsim$ halfway to the ejecta surface.
     This detectability threshold can roughly be reproduced analytically using a two-zone (``kilonova-within-a-supernova'') picture.
     Assuming that a small fraction of SNe produce a detectable $r$-process yield $M_{\rm r} \gtrsim 10^{-2}M_\odot$, and respecting constraints on the total Galactic production rate, we estimate that $\gtrsim 10^{3}-10^4$ SNe need be observed to find one $r$-enriched event, a feat that may become possible with the {\it Vera Rubin Observatory}.

\end{abstract}

\keywords{}

\section{Introduction}

  Current and upcoming large-scale optical time-domain surveys are enabling for the first time the collection of extremely large statistical samples of core-collapse supernova (SN) light-curves.  At present these include facilities such as the {\it Zwicky Transient Facility} ({\it ZTF}; \citealt{Bellm+19}), Asteroid Terrestrial impact Last Alert System ({\it ATLAS}; \citealt{Tonry+18}), Las Cumbres Observatory Global Telescope Network (\textit{LCOGT}; \citealt{Brown2013}) the Automatic Search for Supernovae ({\it ASAS-SN}; \citealt{Kochanek+17}), among others. These efforts will soon be joined by the Vera C.~Rubin Observatory (amongst others, e.g. \citealt{Lin+23}) the first wide-field survey on an 8m-class telescope (\citealt{Ivezic+19}; hereafter {\it Rubin}),  conducting the Legacy Survey of Space and Time (\textit{LSST}).  {\it Rubin} will monitor almost half of the extragalactic sky (roughly 18000 square degrees) over ten years in the $ugrizy$ filters (modeled after the SDSS system), during which it is expected to observe to the order of millions of Type II (hydrogen-rich) SNe \citep{LSST+09}. The Nancy Gray Roman Space Telescope (\textit{Roman}; \citealt{Spergel+15}) is also expected to explore the transient sky with its wide-field infrared capabilities. The resulting light-curve sample will enable detailed studies of the distribution of explosion properties (e.g., \citealt{GB2020,Barker+23}), and how they map onto progenitor star (e.g., \citealt{Podsiadlowski+1992, Crowther2007, Smartt2009, Zapartas+17a, Renzo+19b, Strotjohann+23}) and host galaxy properties (e.g., \citealt{Foley+Mandel2013, Baldeschi+2020, Gagliano+23}).  However, in addition to the expected abundance of ``garden variety'' SNe, such large samples will inevitably reveal {\it anomalous} events (e.g., \citealt{Villar+21}) that offer potential insights into rare but important event classes.

The origin of the rapid neutron-capture process ($r$-process) elements is a long-standing mystery (e.g., \citealt{Horowitz+19,Thielemann+20,Cowan+21}), particularly the astrophysical site or sites which produce the heaviest elements up to the third $r$-process peak around atomic mass number $A \sim 195$.  Core-collapse SNe have long been considered a promising $r$-process source (e.g., \citealt{Meyer+92, Woosley+94, Burrows+95}), motivated in part by observations of metal-poor stars in the Galactic halo or nearby dwarf galaxies which imply an $r$-process site with relatively prompt enrichment after star formation (e.g., \citealt{Ji+16,Cote+19,Simon+23}).  However, the standard neutrino-driven proto-neutron star (PNS) wind believed to accompany many if not all core-collapse SNe, likely fails to produce heavy $r$-process elements (e.g., \citealt{Qian&Woosley96,Otsuki+00,Thompson+01,Wang&Burrows23}).  This has prompted variations of the standard neutrino-wind models, such as convection-driven wave heating (e.g., \citealt{Metzger+07,Nevins&Roberts23}), strong magnetic fields (e.g., \citealt{Thompson03,Thompson+04,Metzger+07,Winteler+12,Mosta+14,Vlasov+17,Thompson&udDoula18,Prasanna+22,Desai+23}) and/or rapid rotation (e.g., \citealt{Desai+22,Prasanna+23}).  In the (likely rare) case that the core of the progenitor is rapidly spinning at the time of the collapse (e.g. \citealt{Ma+Fuller2019}), the latter can lead to an accretion torus forming around the central compact object (e.g., \citealt{MacFadyen&Woosley99}; however, see \citealt{Quataert+Coughlin2019, Burrows+23}, who find disk formation from the collapse of a $40M_{\odot}$ progenitor even without initial rotation).  Such accretion disk outflows may be sufficiently neutron-rich to create $r$-process elements in even larger quantities than possible from the earlier explosion or PNS wind phase (e.g., \citealt{Siegel+19,Miller+20,Just+22}).

A key property of heavy $r$-process elements is the high UV/optical wavelength opacity provided by bound-bound transitions of lanthanide/actinide elements with partially-filled valence electron shells (e.g., \citealt{Kasen+13,Tanaka&Hotokezaka13}). This implies that any portion of the SN ejecta which contains a moderate fraction of these elements will trap radiation more effectively, giving rise to redder emission than an otherwise equivalent $r$-process-free explosion.  Such near-infrared emission is a key diagnostic of heavy element synthesis in the ``kilonovae'' which accompany neutron star mergers (e.g., \citealt{Metzger+10,Barnes_2013, Tanaka+2017}).

Signatures of the $r$-process in kilonovae are conspicuous because neutron star merger ejecta is composed almost exclusively of $r$-process elements.  In contrast, most $r$-process production in a SN explosion will be seeded from the innermost layers and hence expected to be embedded within the bulk of the ``ordinary'' (i.e., non $r$-process-enriched) ejecta originating from the outer layers of the exploding star. Nevertheless, fast-moving $r$-process material as generated in the SN explosion, the neutrino-heated wind, or an accretion disk outflow, can in principle be mixed outwards through the ejecta via hydrodynamic processes (e.g., \citealt{Wongwathanarat+15,Barnes&Duffell23,Wang&Burrows23}), thereby allowing pronounced effects on the SN light-curve even at relatively early times as the SN photosphere retreats through $r$-enriched outer layers.

\citet{Barnes&Metzger22} developed a semi-analytic one-dimensional model for the effects of $r$-process enrichment on the light-curves of stripped-envelope SNe, in which the total mass of $r$-process elements and their degree of radial mixing through the ejecta are treated as free parameters. They found that the effect of $r$-process pollution is to induce reddening in the SN light-curves which becomes more prominent with time, as the photosphere retreats inwards towards the $r$-enriched layers. Motivated by this work, efforts to quantify the presence of $r$-process in stripped-envelope SNe via new or archival infrared follow-up observations have recently begun \citep{Anand+23,Rastinejad+23}. \citet{Barnes&Metzger23} applied their model to interpret the excess infrared emission observed to accompany the nominally long-duration GRB 211211A \citep{Rastinejad+22} to explore whether this event was compatible with being an $r$-process-enriched SN, rather than a kilonova from a neutron-star merger.

In this paper we extend these works to consider the impact that $r$-process enrichment would have on Type IIP SNe (SNe IIP) resulting from the explosions of {\it hydrogen-rich} massive progenitor stars. Most scenarios for substantial $r$-process production in SNe require the birth of a rapidly rotating black hole or neutron star engine (e.g., \citealt{Thompson+04,Metzger+07,Vlasov+17,Prasanna+22,Prasanna+23,Desai+23}), an occurrence less frequently associated with the collapse of hydrogen-rich progenitors than stripped-envelope stars (e.g., \citealt{Kasen&Bildsten10,Piran+17,Metzger+18b}; however, see \citealt{Sukhbold&Thompson17,Dessart+18}).  Nevertheless, with SNe IIP being the most common core-collapse explosion in the universe \citep{Smith+2011}, the large sample of light-curves to be collected over the next years provides an opportunity to search for outlier events, comprising even a rare subset of the sample. Furthermore, as a result of the qualitatively different opacity structure of hydrogen-rich ejecta (namely, electron scattering regulated by hydrogen recombination), $r$-process enrichment can have a distinct effect on the light-curve shape and colors relative to the hydrogen-poor explosions previously studied in \citet{Siegel+19,Barnes&Metzger23}. By developing this framework for SNe IIP, we also lay the groundwork to explore a broader range of SN progenitors in future work.

This paper is organized as follows.  In Sec.~\ref{sec:model} we describe the model for $r$-process-enriched SN light-curves.  In Sec.~\ref{sec:results} we present our results, starting with an illuminating (albeit extreme) fiducial example and then expanding to a broader parameter study. In Sec.~\ref{sec:discussion} we discuss implications of our results, including practical criteria for observationally distinguishing (necessarily rare) $r$-process-enriched explosions from large samples of SNe, and the resulting constraints one can place on the contributions of H-rich SNe to the Galactic $r$-process budget. We summarize our results and conclude in Sec.~\ref{sec:conclusions}.

\section{Model}
\label{sec:model}

We modify the SuperNova Explosion Code\footnote{\url{https://stellarcollapse.org/index.php/SNEC.html}} (\texttt{SNEC}; \citealt{Morozova15}) to account for some of the effects of $r$-process element enrichment on the light-curves of SNe IIP from the explosion of red supergiant (RSG) progenitor stars. \texttt{SNEC} solves the equations of Lagrangian radiation hydrodynamics in one radial dimension under the assumption of equilibrium radiation diffusion with gray opacity. It assumes an equation of state containing contributions from radiation, ions, and electrons \citep{Paczynski1983} with the ionization state calculated from the Saha equations.  As described below in Sec.~\ref{sec:modified}, our main modifications to \texttt{SNEC} include changes to the opacity prescriptions and radioactive heating rate in accordance with what is expected from $r$-process enrichment, parameterized by the total mass of $r$-nuclei, $M_{\rm r}$, and their radial mixing fraction in the SN ejecta, $f_{\rm mix}$.

\subsection{Unenriched Explosion Models}
\label{sec:unenriched}

We consider as a fiducial case the explosion of an (unstripped) nonrotating solar-metallicty ($Z_{\odot}=0.019$) RSG of initial (zero age main sequence) mass $M_{\rm ZAMS} = 15 M_{\odot}$, with a pre-SN mass and radius of $M_{\rm tot} = 12.29 M_{\odot}$ and $R_0 = 1039R_{\odot}$, respectively. The progenitor model accounts for wind mass-loss (\citealt{deJager+1988}, \citealt{Vink+2001}), exponential overshooting, and a 21-isotope nuclear reaction network. It was calculated using the stellar evolution code \texttt{MESA} \citep{Paxton+11,Paxton+13,Paxton+15} and introduced in \cite{renzo:15}; it corresponds to the unstripped model of \cite{Morozova15}.\footnote{The \texttt{MESA} model is included with the \texttt{SNEC} source code.} We first consider a fiducial explosion energy of $E = 10^{51}$ erg. The total mass of the unbound ejecta is $M_{\rm ej} = M_{\rm tot} - M_{\rm NS} = 10.89M_{\odot},$ where $M_{\rm NS} = 1.4M_{\odot}$ is the excised core corresponding to the assumed PNS mass. The explosion is modeled by injecting a large thermal energy approximately equal to $|E_{\rm inital}| + E$ across the inner $0.1 M_{\odot}$ above the PNS mass cut, where $E_{\rm initial}$ is the binding energy of the progenitor at the onset of core collapse. The total mass of radioactive $^{56}$Ni generated during the explosion, $M_{\rm Ni}$, is another free parameter; we assume $M_{\rm Ni} = 0.05M_{\odot}$ in our fiducial model, typical of SNe IIP \citep{Muller2017}. The $^{56}$Ni is assumed to be mixed uniformly from the PNS boundary out to a specified mass coordinate, fiducially taken to be $3M_{\odot}$ (we later apply compositional smoothing to this profile, following the approach of \citealt{Morozova15} as described below), roughly corresponding to the carbon-oxygen core of the progenitor where we would expect most of the explosive $^{56}$Ni nucleosynthesis to occur. \texttt{SNEC} implements artificial boxcar averaging (or ``smoothing") of element compositional profiles to simulate radial mixing from hydrodynamic instabilities arising during shock propagation. We emphasize that \texttt{SNEC} does not include a smoothing mechanism for the density profile (therefore the density profile is inconsistent with the post-boxcar compositions, as seen in Fig.~\ref{fig:initialfiducial}), however, the choice of the minimum opacity profile discussed in the following section attempts to compensate for this.

\subsection{Modified Models Including \textit{r}-Process Enrichment}
\label{sec:modified}

In this study, we do not conduct the detailed multidimensional modeling required to capture the complex physical processes which give rise to neutron-rich ejecta during or following the explosion. Due to the ad-hoc nature of compositional mixing in \texttt{SNEC}, which occurs before the shock passes through the star, and in order to explore the dependence of light-curve properties on the extent of the mixing, we assume the existence of an (early, unmodeled) process that creates $r$-process material near the core of the explosion and radially mixes it into the ejecta in a parameterized way, detailed further below. We modify the continuum opacity and radioactive heating rates in \texttt{SNEC} according to this assumed radial distribution of $r$-process nuclei. The SN light-curve is most sensitive to changes in the opacity.  By contrast, energy deposited from the radioactive decay of $r$-nuclei has only a minor effect on both the ejecta dynamics and the light-curve, as it is typically small compared to the ejecta kinetic energy and energy deposited from $^{56}$Ni decay (as shown in Appendix~\ref{sec:tests}).

In addition to the tabulated temperature and density-dependent opacities from OPAL \citep{Iglesias&Rogers1996} and Ferguson \citep{Ferguson2005}, corresponding to known continuum absorption processes (e.g., bound-free and electron scattering), \texttt{SNEC} implements a minimum opacity ``floor'' proportional to the ejecta metallicity $Z(M)$ in a given mass shell $M$ \citep{Bersten2011, Morozova15},
\be \label{eq: default_floor}
\kappa_{\rm Z}(Z) = \frac{\kappa_{\rm core} Z_\odot - \kappa_{\rm env} + Z(M)(\kappa_{\rm env} - \kappa_{\rm core})}{Z_\odot - 1},
\ee
where we adopt for the solar metallicity $Z_\odot = 0.02$.
In the outer hydrogen-rich layers of the ejecta, the floor opacity converges to a default ``envelope'' value $\kappa_{\rm Z}(M_{\rm ej} + M_{\rm NS}) = \kappa_{\rm env} = 0.01~\mathrm{cm^2 g^{-1}}$ roughly corresponding to the metal line opacity of solar-composition material. For mass coordinates approaching the innermost ejecta near the core boundary, the default floor opacity is higher, taken to be $\kappa_{\rm Z}(M_{\rm NS}) = \kappa_{\rm core} = 0.24~\mathrm{cm^2 g^{-1}}$ corresponding to the higher metallicity of these layers. Although the above values for $\kappa_{\rm env}$ and $\kappa_{\rm core}$ may be higher than those expected physically for ordinary SN composition material, they are chosen for \texttt{SNEC }to produce SN IIP light-curve shapes in agreement with light-curves from multi-group calculations \citep{Bersten2011, Bersten2013} and observations \citep[e.g.,][]{Morozova15}. In the context of considering even higher opacities arising from $r$-process elements, this relatively high floor opacity for our unenriched models is conservative in that it tends to mitigate the effects of $r$-process enrichment. In Appendix~\ref{app:defaultfloor} we also show light-curves calculated for a floor opacity reduced by an order of magnitude, finding that the relatively high opacity floor prescription cited above is required to avoid unphysical light-curve features, in part arising due to the unsmoothed density profile.

To the extent that the pseudo-continuum line opacity of $r$-process elements (e.g., \citealt{Kasen+13}) can be approximately modeled as a gray opacity roughly independent of temperature \citep{Tanaka+20}, we model their presence in the ejecta by modifying the opacity floor profile in \texttt{SNEC}, according to:
\be \label{eq:kappa_f}
\kappa_{\rm f}(M) = \kappa_{\rm r}(X_{\rm r}) + \kappa_{\rm Z}(Z).
\ee
where $\kappa_{\rm Z}$ is the normal opacity floor of Eq.~\eqref{eq: default_floor}. The opacity contribution at coordinate $M$ of $r$-process elements, $\kappa_{\rm r}$, is taken to be proportional to the $r$-process mass-fraction, $X_{\rm r}(M)$, as follows:
\be
\kappa_{\rm r} = \kappa_{\rm rp}X_{\rm r}(M),
\label{eq:kappar}
\ee
where $\kappa_{\rm rp}$ is the effective gray opacity of pure $r$-process material. We expect a range of values $\kappa_{\rm rp} \approx 3-30$ cm$^{2}$ g$^{-1}$, depending on the abundance distribution of $r$-nuclei, particularly the mass fraction of lanthanide/actinide elements (e.g., \citealt{Kasen+13,Tanaka&Hotokezaka13}). Hereafter, we take $\kappa_{\rm rp} = 20$ cm$^{2}$ g$^{-1}$, corresponding roughly to the expected value for material with a solar $r$-process abundance ratios at a temperature $T \approx 5000$ K (e.g., \citealt{Tanaka+20}).

Lacking a first principles model for the radial distribution of $r$-process elements within the ejecta, we adopt a somewhat ad hoc mass-fraction profile:
\be \label{eq:Xr}
X_{\rm r}(M) = \mathcal{C}\left[\tanh\left(\frac{M_{\rm r,mix}-M}{\chi M_{\rm r,mix}}\right)+1\right].
\ee
Here, $M_{\rm r,mix}$ is the characteristic mass coordinate out to which $r$-nuclei are homogeneously mixed (e.g., during the SN explosion itself or due to delayed outflows from the central compact object$-$magnetar or accreting black hole) and $\chi$ is a dimensionless parameter proportional to the radial width of the transition between $r$-process-enriched and unenriched layers.  We take $\chi = 0.2$, in order to give a relatively sharp transition from $r$-enriched layers at $M \ll M_{\rm r,mix}$ to $r$-process-free layers at $M \gg M_{\rm r,mix}$ (see \citealt{Barnes&Duffell23} for motivation), though our results are not overly sensitive to this precise value, as long as the transition remains sharp (as shown in Appendix~\ref{app:chi_effects}).  Rather than $M_{\rm r,mix}$, it will often be more convenient to quote the ejecta radial mixing fraction,
\be \label{fmix}
f_{\rm mix} \equiv \frac{M_{\rm r,mix} - M_{\rm NS}}{M_{\rm ej}},
\ee
which we take as a free parameter. Small $f_{\rm mix}$ represents $r$-process that remains confined to the center of the star and larger $f_{\rm mix}$ approaching unity represent mixing that extends all the way to the ejecta surface. While the radial mixing extent of $r$-process rich outflows has not been studied directly, 3D simulations and observation support the possibility of metal-rich clumps of ejecta (e.g., ``nickel bullets'' facilitated by Rayleigh-Taylor instabilities) penetrating into the H-envelope (\citealt{Spyr+1990,Fassia+1999,Wongwathanarat+15,Sandoval+21}). Given the open question of whether $r$-process can mix in a similar fashion, in this work, we explore the full range of possible mixing extents $f_{\rm mix} = 0.1 - 1.0$.

The normalization constant $\mathcal{C}$ entering Eq.~\eqref{eq:Xr} is set by the total mass of $r$-process nuclei, $M_{\rm r} = \int X_{\rm r}dM,$ which is also treated as a free parameter.  The expected $r$-process yields of the explosion can vary considerably, depending on the nature of the central engine responsible for generating neutron-rich $r$-process material.  For example, the neutrino-driven winds that accompany PNS birth in most core-collapse SNe yield $M_{\rm r} \approx 10^{-5}-10^{-4}M_{\odot}$ (e.g., \citealt{Qian&Woosley96}) while the prodigious collapsar-like accretion disk outflows can conceivably reach $M_{\rm r} \gtrsim 1 M_{\odot}$ or even higher for very massive stars (\citealt{Siegel+19,Siegel+22}; though we note that other simulations making different assumptions regarding the transport of angular momentum and neutrinos in the disk differ on whether heavy $r$-process nuclei are generated in collapsar disks; e.g., \citealt{Miller+20,Fujibayashi+20,Just+22}).

The second effect of $r$-process enrichment is on the radioactive heating rate within the SN ejecta. In addition to the usual heating from the $^{56}$Ni $\rightarrow ^{56}$Co$\rightarrow ^{56}$Fe decay chain already implemented in \texttt{SNEC}, we add a specific radioactive heating rate for $r$-process nuclei of the form \citep{Metzger+10}
\be \label{eq:epsilon_rp}
\epsilon_{\rm rp} = 3\times 10^{10}\,\mathrm{erg\,g^{-1}\,s^{-1}}\left(\frac{t}{1\,{\rm day}}\right)^{-1.3},
\ee
where $t$ is time measured since the explosion.  The chosen normalization of the heating rate is on the high end of the $\beta$-decay heating rate for solar-like $r$-process distribution (e.g., \citealt{Wanajo18}) and we neglect corrections due to inefficient thermalization of the radioactive decay products (e.g., \citealt{Barnes_etal_2016}).  Even under such generous assumptions, we show in Appendix~\ref{app:heatingeffects} that the heating effects, even for large $r$-process masses, are generally minor. This follows because the heating rate of the $^{56}$Ni $\rightarrow ^{56}$Co$\rightarrow ^{56}$Fe decay chain greatly exceeds the $r$-process heating rate on the weeks to months timescales relevant to SN light-curves \citep{Siegel+19,Barnes&Metzger22}, and because the high opacity of $r$-enriched ejecta efficiently traps the thermalized energy released earlier from $r$-process decay, diluting its effects due to adiabatic expansion.

In summary, for a given stellar progenitor, an $r$-process-enriched model is fully specified in our setup by the free parameters $\{M_{\rm r}, f_{\rm mix}\}$, with $\kappa_{\rm rp}$ and $\chi$ fixed at the aforementioned values.

\section{Results}
\label{sec:results}
\subsection{Fiducial Model} \label{sec:fiducial}

\begin{figure*}
    \centering
    \includegraphics[width=1.0\textwidth]{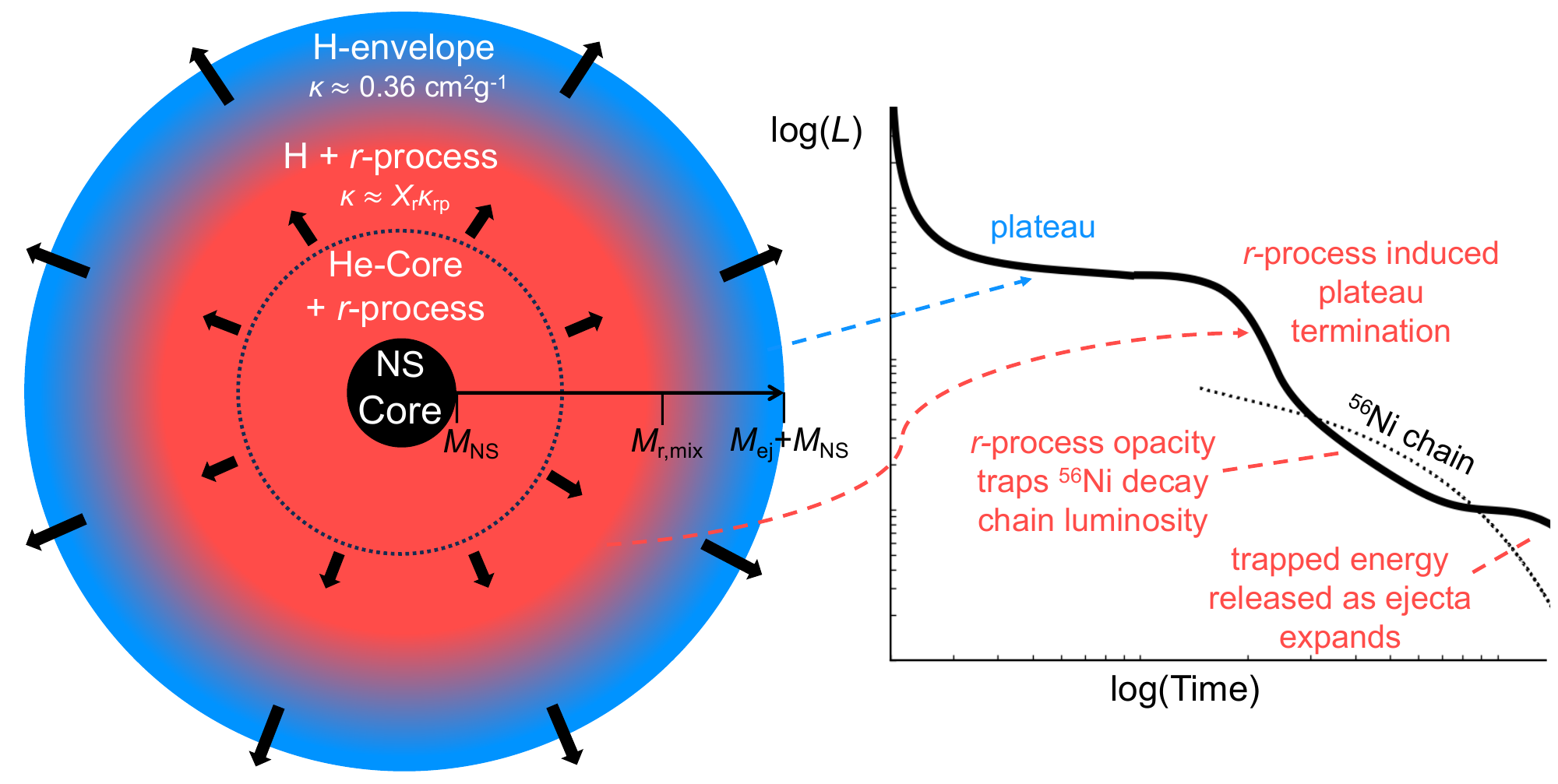}
    \caption{\normalsize Schematic illustration of the effects of a centrally-concentrated source of $r$-process elements on the ejecta structure of hydrogen-rich SNe (left) and how the inwards retreat of the photosphere through these layers maps into phases of the light-curve (right).}
    \label{fig:cartoon}
\end{figure*}

We first describe the results for a fiducial, highly $r$-process-enriched model with $M_{\rm r} = 0.3 M_{\odot}$ and $f_{\rm mix} = 0.7$, the latter corresponding to a case where $r$-process material is mixed out through the ejecta to a mass-coordinate $M_{\rm r,mix} = f_{\rm mix}M_{\rm ej} + M_{\rm NS} \approx 9M_{\odot}$, located below the ejecta surface but within the hydrogen-rich envelope for the fiducial $M_{\rm tot} = 12.29 M_\odot$. Although $M_{\rm r} = 0.3 M_{\odot}$ represents a very large $r$-process yield, this model is instructive for illustrating clearly the effects of $r$-enrichment.  Throughout, we compare the model evolution side by side to an otherwise equivalent but ``unenriched'' $r$-process-free model with $M_{\rm r} = 0$, which successfully reproduces the fiducial model of \cite{Morozova15}.  Fig.~\ref{fig:cartoon} summarizes schematically the physical picture.

\begin{figure}
    \centering
\includegraphics[width=0.5\textwidth]{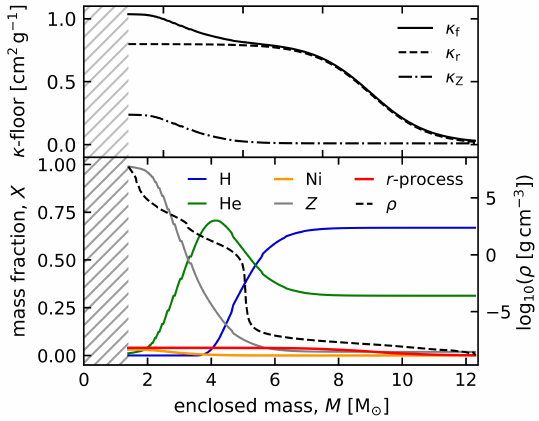}  
    \caption{\normalsize Ejecta composition and radial density profiles (bottom panel) and opacity floor profiles (top panel) as a function of enclosed mass, for the fiducial highly $r$-process-enriched model with $\{M_{\rm r} = 0.3 M_{\odot}, \ f_{\rm mix} = 0.7\}$. The opacity floor of an enriched model is the sum of the ``default" floor of the unenriched model and an $r$-process contribution which tracks the assumed $r$-process mass profile. A shaded region marks the excised PNS core.}
    \label{fig:initialfiducial}
\end{figure}

Fig.~\ref{fig:initialfiducial} shows the opacity floor (top panel), along with the mass fractions of different elements and density as a function of enclosed mass (bottom panel). We include in the total opacity floor, $\kappa_{\rm f}$ (Eq.~\eqref{eq:kappa_f}), the separate contributions from the $r$-process elements, $\kappa_{\rm r} \propto X_r(M)$, and the floor set by ordinary non $r$-process material, $\kappa_{\rm Z} \propto Z(M)$. The former falls off from its uniform value near the core to zero for $M > M_{\rm r,mix} \approx 9M_{\odot}.$ Due to the high opacity contributed by the $r$-process nuclei, the opacity floor at small radii exceeds that of the unenriched model by a factor $\approx 3$. Note, the metallicity \textit{Z} does not include the additional $r$-process elements, which are included in the opacity calculation via the modified opacity floor and are otherwise treated as a perturbation to the total ejecta mass and metal distribution. Furthermore, the smoothing assumption does not apply to the density, thus the steep transition at $M \approx 5 M_{\odot}$.

\begin{figure}
    \centering
    \includegraphics[width=0.5\textwidth]{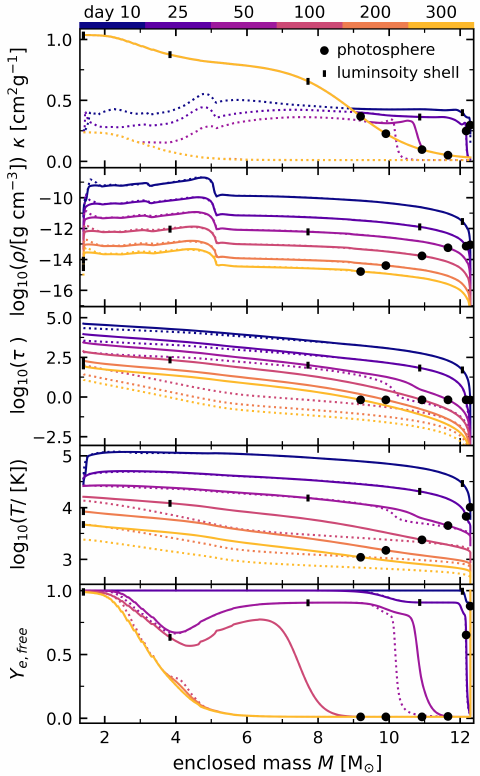}
    \caption{\normalsize Radial profiles for the fiducial $r$-process-enriched model $\{M_{\rm r} = 0.3 M_{\odot}$, $f_{\rm mix} = 0.7\}$. Solid lines show (from top to bottom) opacity, density, optical depth coordinate, temperature, and free electron fraction, respectively, at various times ranging from 10 to 300 days after explosion as labeled. Dotted lines show the profiles for an otherwise equivalent ``unenriched'' $r$-process-free model ($M_{\rm r} = 0$).  Filled circles and vertical tick marks show the location of the photosphere and the ``luminosity shell" (defined as where the optical depth $\tau \approx c/v$), respectively, in the fiducial enriched model. The sharp drop in the density at $M \approx 5 M_\odot$ is due to the assumed unsmoothed initial density profile (Fig.~\ref{fig:initialfiducial}).}
    \label{fig:timelapse}
\end{figure}

Fig.~\ref{fig:timelapse} shows snapshots of the radial profiles of several key ejecta properties at different times ranging from 10 to 300 days after explosion.  The early phases of the explosion, when the photosphere is still in the outer $r$-process-free layers, proceed in a similar manner to an ordinary SN IIP from an $r$-process-free explosion (e.g., \citealt{Grassberg+1971,Falk&Arnett1977,Litvinova&Nadezhin1985,Popov93}). The radial density profile is determined by the progenitor structure and the SN shock that passes through the ejecta within the first 48 hours following the explosion and its subsequent approach to a state of homologous expansion. As the ejecta continues to expand, its temperature at all radii drops monotonically with time due to a combination of adiabatic and radiative losses.  As the outer layers reach the recombination temperature, $T_{\rm rec} \approx$ 6000 K, the free electron fraction there decreases from near unity to almost zero, causing an abrupt drop in the opacity from the electron scattering value $\kappa \simeq \kappa_{\rm es} \approx 0.3$ cm$^{2}\,$g$^{-1}$ to the lower floor opacity of the envelope. The location of the photosphere is therefore fixed at the location where $T_{\rm rec} \approx$ 6000 K, which retreats inwards in mass-coordinate with time, driving the characteristic ``plateau" phase in SNe IIP. The opacity evolution described above for mass coordinates $M\gtrsim M_{\rm r,mix} \approx 9M_{\odot}$ can be seen in top panel of Fig.~\ref{fig:timelapse} over the first two snapshots of the enriched model corresponding to $t \lesssim 25$ d.

However, by $t \approx 30$ d, the photosphere has retreated sufficiently to reveal zones in which the $r$-process opacity exceeds alternate opacity sources (e.g. electron scattering or unenriched floor). From this time on, the opacity is dominated by the time-independent floor value for all mass coordinates, which for the remaining optically-thick layers now corresponds to the opacity $\kappa \approx \kappa_{\rm f} \gtrsim 0.5$ cm$^{2}\,$g$^{-1}$ arising primarily from $r$-process elements. This high and (at least within our treatment) temperature-independent opacity thus comes to dominate the ejecta cooling after the first month, altering the emission behavior from that of the unenriched model as we now describe.

\begin{figure}
    \centering
    \includegraphics[width=0.5\textwidth]{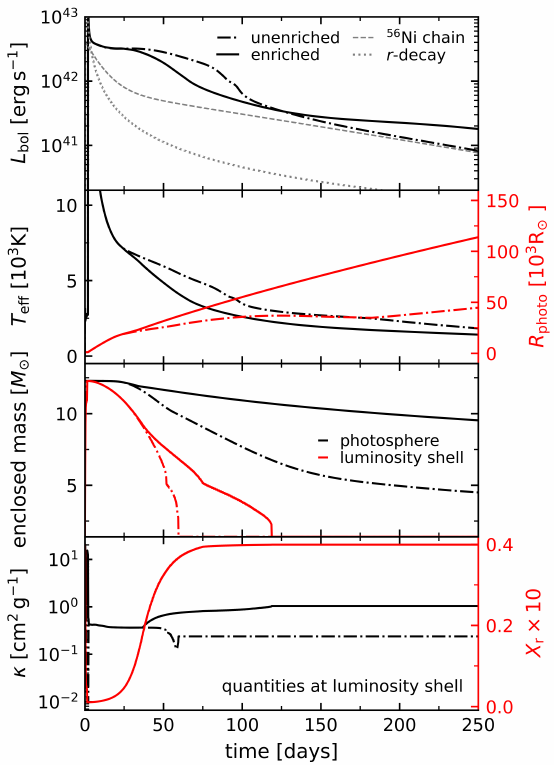}
    \caption{\normalsize Comparison of the time evolution of various quantities for the fiducial $r$-process-enriched (solid lines) and unenriched (dot-dashed lines) models, respectively. From top to bottom we show: bolometric light-curve (with the total radioactive heating rate from the $^{56}$Ni chain and $r$-process elements shown for comparison); effective temperature (left vertical axis, black) and photosphere radius (right vertical axis, red); mass-coordinates of the photosphere (black) and luminosity shell (red); and opacity value (left vertical axis, black) and $r$-process mass fraction (right vertical axis, red) of the luminosity shell.}
    \label{fig:fiducial}
\end{figure}

The top two panels of Fig.~\ref{fig:fiducial} show the bolometric light-curve and effective temperature, respectively, for the fiducial $r$-process-enriched and unenriched models. In ordinary SNe IIP as illustrated by the unenriched model, the early shock-cooling phase is followed by the characteristic plateau phase of roughly constant luminosity lasting approximately 100 days with a roughly constant effective temperature $T_{\rm eff} \approx T_{\rm rec} \approx 5000-6000$ K as determined by the hydrogen recombination front regulating the photosphere location.  After the photosphere has moved completely through the H-envelope and into the He-rich core, the plateau ends and the luminosity drops, until it eventually converges on the energy output from the $^{56}\rm Ni$ decay chain.

However, the luminosity and temperature for the enriched model begin to deviate significantly from the unenriched model once the photosphere reaches the $r$-enriched layers around day 30. In particular, the length of the plateau phase, $t_{\rm p}$, is shortened for the enriched model by $\approx$30 days compared to the unenriched case. The magnitude of this shortening is consistent with when the opacity at the luminosity shell (the location in the ejecta where $\tau = c/v$, beyond which radiation can stream to the observer within a dynamical time; bottom panel of Fig.~\ref{fig:fiducial}) rises from the electron scattering to $r$-process opacities around day 30 (see also Fig.~\ref{fig:timelapse}). After the plateau ends, the photosphere location is no longer set by the hydrogen recombination front, despite the fact that not all of the hydrogen has recombined.  Instead, after day 30 the photosphere stalls at a higher mass coordinate due to the high $r$-process opacity in what are still (by mass) H-dominated layers, as shown in the third panel of Fig.~\ref{fig:fiducial}.

The lower effective temperature in the enriched case results from the higher opacity pushing the photosphere radius $R_{\rm photo}$ out into the cooler outer ejecta layers, as can be seen by comparing the second panel of Fig.~\ref{fig:fiducial} and the fourth panel of Fig.~\ref{fig:timelapse}. This larger $R_{\rm photo}$ is not sufficient to compensate for the lower temperature in setting the bolometric luminosity $L \propto R_{\rm photo}^2 T_{\rm eff}^4$, resulting in the enriched model light-curve decaying faster than the unenriched case. However, at very late times $t \gtrsim 200$ d, the enriched model becomes brighter than the unenriched model; this results from the high $r$-process opacity in the slowly expanding core material, which traps the $^{56}$Ni chain radioactive decay and initial thermal energy longer than in the unenriched case to be released at these later times.

In summary, the main effects of significant $r$-process enrichment extending into the hydrogen envelope on the SN properties are: (1) a premature truncation of the light-curve plateau phase once the H-recombination photosphere enters the $r$-process-enriched layers, after which follows (2) a more rapid decline in the luminosity and a lower effective temperature at a fixed time after the explosion leading to a premature reddening of the SN colors; (3) a late luminosity in excess of the $^{56}$Ni decay-chain tail resulting from heat trapped by the high opacity of the $r$-enriched core material.

From the above trends, we expect the timescale of plateau shortening to correlate primarily with the mass coordinate of the $r$-process mixing $M_{\rm r,mix}$ (equivalently, $f_{\rm mix}$), while the magnitude of the deviations to follow the plateau will scale with the level of the enrichment $M_{\rm r}$, as this controls the opacity enhancement of the inner layers which radiate at late times. These possibilities are explored in Sec.~\ref{sec:paramstudy}.

\subsection{Parameter Study}
\label{sec:paramstudy}

\begin{figure*}
    \centering
    \includegraphics[width=1.0\textwidth]{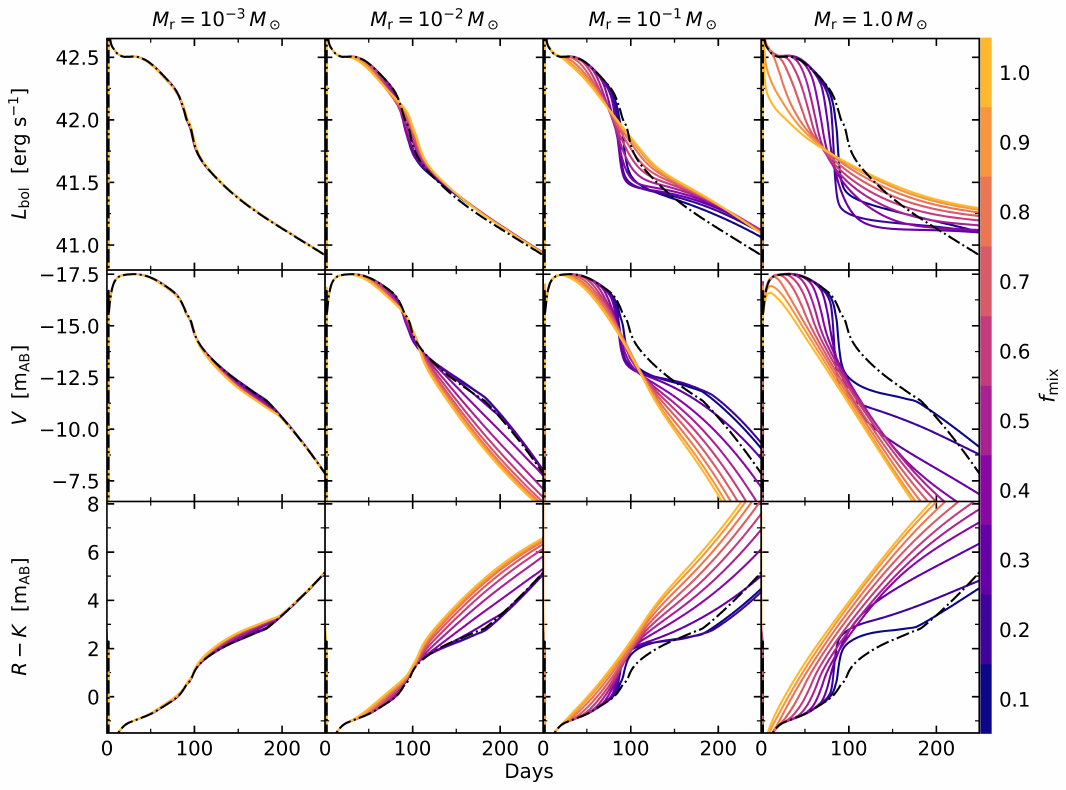}
    \caption{\normalsize Bolometric (top panels) and $V$-band light-curves (middle panels), and $R-K$ color index (bottom panels), for a range of assumptions for the $r$-process mass yield $M_{\rm r}$ (columns, as marked along the top) and the radial mixing fraction of $r$-process material $f_{\rm mix}$ (color scheme). AB magnitudes are assumed, calculated from the bolometric luminosity assuming a  blackbody spectrum. A black dot-dashed line shows for comparison an unenriched but otherwise identical model ($M_{\rm r} = 0$).}
    \label{fig:grid}
\end{figure*}

We now explore a broader range of light-curve behavior for models spanning different $r$-process ejecta masses $M_{\rm r} = 10^{-3} - 1 M_{\odot}$ and radial mixing extents $f_{\rm mix} = 0.1 -1$. Fig.~\ref{fig:grid} shows for such a grid of models, the bolometric and $V$-band light-curves, as well as the $R-K$ color evolution (motivated by \citealt{Barnes&Metzger22}, who found $R-K$ is a sensitive diagnostic for $r$-enrichment).\footnote{See Appendix~\ref{app:LSSTfig} for an alternate version of Fig.~\ref{fig:grid} with \textit{LSST} bands.} Photometric band AB magnitudes are calculated in post-processing from the bolometric luminosity assuming a blackbody spectrum.

The models with the lowest levels of $r$-process enrichment (e.g., $M_{\rm r} = 10^{-3} M_{\odot}$) show bolometric light-curves virtually indistinguishable from an otherwise identical unenriched model, and the effects in $V$-band and $R-K$ color evolution are also small. This indicates that the prospect of photometric detection for $M_{\rm r} \ll 10^{-2} M_{\odot}$ are limited. However, for models with larger $M_{\rm r} \gtrsim 10^{-2} M_{\odot}$, the three major effects of $r$-process enrichment, described in Sec.~\ref{sec:fiducial}, become apparent.

The degree of plateau shortening increases as $r$-process material mixes further out in the ejecta (larger $f_{\rm mix}$) due to the high-opacity regions being exposed earlier, evident in the bolometric and $V$-band light-curves of Fig.~\ref{fig:grid}. Moreover, for models with $f_{\rm mix} \gtrsim 0.4$ for which plateau shortening is visible, greater $r$-process mass amplifies the effect. For example, although the bolometric plateau duration $t_{\rm p}$ for the $M_{\rm r} = 0.1M_{\odot}$ and $M_{\rm r} = 1.0M_{\odot}$ cases are equal to within a few days for $f_{\rm mix} = 0.1$, the difference becomes much more pronounced for $f_{\rm mix} = 1$ ($t_{\rm p} \approx 50-60$ days for $M_{\rm r} = 0.1M_{\odot}$ versus a complete absence of a plateau for $M_{\rm r} = 1 M_{\odot}$).

A second effect of $r$-process enrichment is to reduce the luminosity and effective temperature after the plateau phase. Regarding this declining luminosity phase, there are two general behaviors, depending on $f_{\rm mix}$. When $r$-process material is concentrated at the center of the ejecta (i.e., for low $f_{\rm mix}$), the plateau is followed by an abrupt and steep drop in luminosity; in extreme cases (e.g., $M_{\rm r} \gtrsim 0.1 M_{\odot}$ with $f_{\rm mix} = 0.1$) the luminosity drops by an order of magnitude or more within days after the plateau ends. Conversely, models with greater $f_{\rm mix}$ (e.g., $M_{\rm r} \gtrsim 0.1 M_{\odot}$ with $f_{\rm mix} = 1.0$), associated with shorter plateaus, experience a linear or almost-linear decline in both bolometric and $V$-band luminosity. As an aside, we note that this (probably coincidentally) mimics the behavior of some SNe IIL light-curves, which are normally attributed to the SN progenitor being partially stripped of its H-envelope \citep[e.g.][]{Hiramatsu2021b}, and/or reheating of the ejecta by circumstellar interactions (in certain parameter regimes of high explosion energies and/or low $^{56}$Ni mass; \citealt{Morozova15,Moriya+2016,Morozova_2017,Fraser20,Piro+21}).

These behaviors are physically consistent with expectations given the enriched ejecta structure. When the $r$-process mass is significant ($M_{\rm r} \gtrsim 0.1 M_{\odot}$) but confined to small mass coordinates ($f_{\rm mix} \approx 0.1$), the recombination driven phase will proceed as normal at early times, until the photosphere recedes sufficiently far in to hit an ``opacity wall'' of a highly concentrated $r$-process core at $\approx M_{\rm r,mix}$, after which the photosphere abruptly stalls, leading to an abrupt drop in luminosity. On the other hand, when a significant quantity of $r$-process material is mixed more uniformly throughout the ejecta ($f_{\rm mix} \approx 1$), no such abrupt transition from low to high opacity layers occurs, preventing sudden changes in the light-curve decay. Instead, the escaping luminosity even from early phases is inhibited by the $r$-process opacity effects, resulting in a comparatively steady decline rate.

Models with greater $r$-process enrichment mixed to larger mixing coordinates are in general also significantly redder than models with lower enrichment at a given time, as shown in the bottom row of Fig.~\ref{fig:grid}. Consistent with our explanation for the photosphere evolution in the fiducial model (Sec.~\ref{sec:fiducial}), this trend of redder emission with increasing $M_{\rm r}$ and $f_{\rm mix}$ is due to a slower photosphere recession rate through the outer ejecta layers, resulting in a lower $T_{\rm eff}$. Models with extremely low $f_{\rm mix}$ are an exception to this, insofar that their $R-K$ colors become even redder than more homogeneously mixed models around the time of the abrupt steep luminosity drop.
This results from the photosphere radius stalling or even reversing direction (in radial coordinate) once reaching the highly-concentrated $r$-process core, leading to a significant drop in the effective temperature.

In summary, considering a wide parameter space of $r$-process enrichment profiles $\{ M_{\rm r}, f_{\rm mix} \}$, noticeable differences appear in the light-curves for $M_{\rm r} \gtrsim 10^{-2} M_{\odot}$. The light-curve behavior for models which satisfy this minimal condition are further distinguished based on the radial mixing parameter. Models with low $f_{\rm mix}\approx0.1$ exhibit less plateau shortening, but a steep transition to the nebular phase (defined as when the ejecta becomes optically thin, rendering the innermost layers visible), while models with large $f_{\rm mix}$ exhibit more severe plateau shortening and a corresponding roughly linear light-curve decline. These features, for both low and high degrees of mixing, become more apparent with greater $r$-process mass. Likewise, all models with $M_{\rm r} \gtrsim 10^{-2}M_{\odot}$ exhibit noticeable reddening, although the exact color evolution differs based on $f_{\rm mix}$ in a manner consistent with the luminosity behavior. As in the fiducial model, the very late-time luminosity for all sufficiently enriched models is enhanced by $r$-process opacity trapping, the magnitude of the effect scaling with $M_{\rm r}$.

\section{Applications and Discussion}
\label{sec:discussion}

\subsection{Detectability of \textit{r}-Process Enrichment}
\label{sec:detectability}

 In the previous section we found that SNe enriched in $r$-process masses $M_{\rm r} \gtrsim 10^{-2}M_{\odot}$ exhibit sizable photometric differences relative to otherwise similar unenriched explosions (Fig.~\ref{fig:grid}). We now attempt to more precisely quantify the ``detectability'' prospects for $r$-process-enriched events by considering several photometric and spectroscopic observables commonly used in the transient community to characterize observed SNe which correlate with varied explosion, progenitor star, and host galaxy properties. We choose five such metrics that we found to be most sensitive to $r$-process enrichment:

\begin{itemize}
    \item $t_{\rm p}$: the plateau duration.\footnote{Obtained via a fitting procedure based on \citet{Valenti+2016} and also adopted by \citet{Goldberg+19}; see Appendix~\ref{app: fitting} for details).}

    \item $L_{50}$: the bolometric luminosity at day 50.

    \item $v_{50}$: the ejecta velocity at the photosphere location at day 50.

    \item $s$: the absolute value of the plateau slope in the $V$-band light-curve, analogous to the $s_{\rm 2}$ metric defined in \citet{Anderson+14}.\footnote{These values are obtained by a linear fitting procedure described in Appendix~\ref{app: fitting}.}

    \item $(R-K)_{100}$: The $R-K$ color at day 100.

\end{itemize}

\begin{figure*}
    \centering\includegraphics[width=1.0\textwidth]{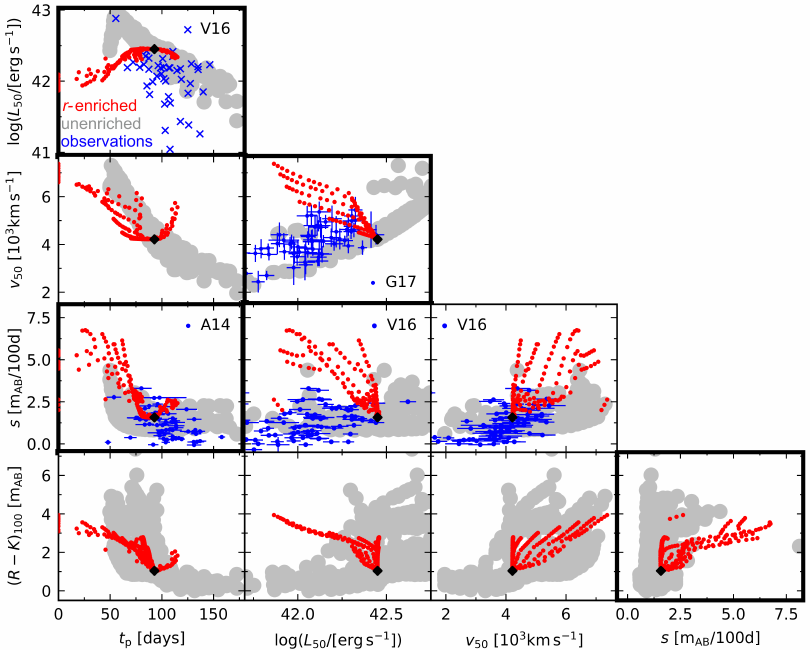}
    \caption{\normalsize Parameter space of observable metrics and values extracted from our light-curve models: $t_{\rm p}, L_{50}, v_{50}, s$, and $(R-K)_{100}$ (see text for definitions), which highlight the locations occupied by $r$-process-enriched explosions. The fiducial enriched model (Sec.~\ref{sec:fiducial}) is shown with a black diamond, while red circles show for the same progenitor model different levels of $r$-process enrichment ($M_{\rm r} = 10^{-3} - 1 M_{\odot}; f_{\rm mix} = 0.1 - 1.0$; see Fig.~\ref{fig:grid}). Large gray circles show unenriched ($M_{\rm r} = 0$) models, sampling different explosion energies ($E = 10^{\rm 50} - 10^{\rm 52} \rm erg $), $^{56}$Ni masses ($M_{\rm Ni} = 0.001 - 0.1 M_{\odot}$) and stellar progenitor models ($M_{\rm tot} = 10 - 13.2 M_{\odot}$). Also shown for comparison with blue points are metrics obtained directly from SNe IIP observations, as reported in \citet[][A14]{Anderson+14}, \citet[][G17]{Gutierrez+17}, and \citet[][V16]{Valenti+2016}. The four plots outlined in bold, $L_{50} \text{ vs. } t_{\rm p}$, $ v_{50} \text{ vs. } L_{50}$, $s \text{ vs. } t_{\rm p}$, $\text{ and}$, $(R-K)_{100} \text{ vs. } s$ are those we find most cleanly enable the identification of enriched models, and for which we define the $r$-process ``detectability'' summary metric used to create Fig.~\ref{fig:detectability}.}
    \label{fig:corner}
\end{figure*}

All possible 2D covariant parameter spaces obtained from these five metrics are shown in Fig.~\ref{fig:corner}. Each space is populated by enriched models spanning parameter values $\{ M_{\rm r} = 10^{-4} - 1 M_{\odot} \}$ and $\{ f_{\rm mix} = 0.1 - 1.0 \}$. These are compared to a large sample of unenriched models, spanning different $^{56}$Ni masses $\{ M_{\rm Ni} = 0.001 - 0.1 M_{\odot} \}$, explosion energies $\{ E = 10^{50} - 10^{\rm 52}\,\rm{erg} \}$, and assumed stellar progenitor models\footnote{\texttt{MESA} progenitor models from \citet{Morozova+2016} available at \url{https://stellarcollapse.org/Morozova2016.html}. See also \cite{renzo:15}.}  (\citealt{Morozova+2016}) covering a range of pre-explosion masses $\{ M_{\rm tot} = 10 - 12.6 M_{\odot} \}$. When available, we also show for comparison measured values for these metrics obtained from SNe IIP observations collected from the literature. While sufficiently enriched (i.e., high $M_{\rm r}$ and/or high $f_{\rm mix}$) models occupy unique regions in all parameter spaces, we focus below on some cases where this separation between enriched and unenriched models is the cleanest (these parameter spaces are outlined in bold in Fig.~\ref{fig:grid}).

Quantities at day 50 are of interest in ordinary SNe IIP as they typically sample the plateau phase. For instance, $L_{50}$ and $t_p$ can probe the explosion energy and properties of the progenitor star (\citealt{Popov93}, \citealt{Kasen&Woosley2009}, though see discussions in \citealt{Dessart+19} and \citealt{Goldberg+19} regarding the non-uniqueness of such constraints due to variations in progenitor radius and H-rich envelope mass). In general, weaker explosions are fainter, and, due to slower expansion velocities, exhibit longer plateaus. This anti-correlation between $L_{50}$ and $t_{\rm p}$ is revealed in the data and reproduced by our unenriched models. Conversely, larger $r$-process enrichment drives shorter plateaus and reduced effective temperatures, resulting in a dimmer plateau luminosity. For very high levels of enrichment (severely shortened plateaus), it is sometimes the case that the plateau has already terminated by day 50, in which case $L_{50}$ samples the subsequent post-plateau decay phase, breaking the general trends. Nevertheless, the enriched model sequence follows a distinct track compared to the unenriched models, remaining clearly separated in $L_{50}-t_{\rm p}$ space.

A second feature of significance is the deviation of $r$-process-enriched SNe from the well-documented SN IIP ``standard candle" relationship between the photospheric velocity $v_{\rm 50}$ and luminosity $L_{\rm 50}$ at day 50 (as observed by \citealt{Hamuy2003} and explained by \citealt{Kasen_2009}).
The photosphere velocity is commonly inferred from the absorption minimum of the Fe II $\lambda$5169\AA\ line, which corresponds to the ejecta velocity near the photosphere. Since the photosphere location in an unenriched SN IIP is set by the recombination of H at fixed $T_{\rm ph}\approx T_{\rm I}\approx6000$K, unenriched SNe IIP have been shown to follow a relationship $L_{50}\appropto v_{\rm 50}^2$ (see, e.g., discussions by \citealt{Kasen_2009,Goldberg+19}). This relation is evident in both the data and unenriched models of Fig.~\ref{fig:corner}.

In contrast, the $r$-process-enriched models exhibit an {\it anticorrelation} between $L_{50}$ and $v_{50}$, with higher levels of enrichment pushing these events into a parameter space region unoccupied by any unenenriched models or SN data. This behavior can be attributed to the photosphere's recession in mass coordinate being stalled due to the high $r$-process opacity as detailed in Sec.~\ref{sec:fiducial}. This keeps the photosphere out at larger radii where the ejecta expansion velocity is greater (homologous expansion) and the temperature is lower.
That is, unlike in an unenriched SN IIP, the photosphere location is not set by the location where the ejecta temperature equals the H-recombination temperature, but rather by the changing density of the expanding medium for a given magnitude and location of the large $\kappa_{\rm rp}$ (which is taken here to be temperature- and density-independent).
The anti-correlation can thus approximately be understood
considering homologous expansion ($R \approx v t$ for a shell of ejecta moving at velocity $v$) and the Stephan-Boltzmann
equation $L(t)\approx 4\pi R_{\rm ph}^2 \sigma_{\rm SB} T_{\rm ph}^4$
(up to a dilution factor which is assumed to primarily be a function of $L$, see discussion by \citealt{Kasen_2009}).
When $T_{\rm}$ is set by $T_{\rm I}\approx6000$K, and time $t$ is fixed at $50$ days, this leads to the familiar $L_{50}\appropto v_{50}^2$ for ordinary SNe.
If the photospheric temperature is instead set by the ejecta's expansion, e.g. $T_{\rm ph}\appropto 1/R_{\rm ph}$ (or any power-law dependence steeper than $T_{\rm ph}\appropto R_{\rm ph}^{-1/2}$), then for a fixed time
we would instead see an anti-correlation $L_{50}\appropto v_{50}^{-2}$ (or a similar inverse relationship) with scatter introduced from varying $M_r$ and $f_{\rm mix}$.
The opposing trends between unenriched and enriched models support $v_{50}$ vs. $L_{50}$ as a promising parameter space in which to flag potential $r$-process sources within the first 50 days of SN evolution.

The $R-K$ color evolution was shown to be a sensitive metric for $r$-enrichment in stripped-envelope SNe (\citealt{Barnes&Metzger22}) and we show this is also the case for hydrogen-rich explosions (Sec.~\ref{sec:paramstudy}). We consider the color value at day 100, as this timescale is typically sufficient to develop appreciable differences from the unenriched case. \citet{Patat+94} showed that SNe IIL, arising from the core-collapse of partially stripped envelope progenitors, or IIL-like explosions possess steeper plateaus (larger decline rates, $s$) along with reduced reddening at day 100. We show that $r$-process-enriched models exhibit a unique correlation in $(R-K)_{100}$ vs. $s$ space, where increasing enrichment results in {\it both} a steeper decline rate and increased reddening. The extreme decline rates of our highly enriched models place them in a unique region of this parameter space.

As was discussed in Sec.~\ref{sec:paramstudy}, light-curves for $r$-process-enriched models with a large degree of radial mixing exhibit a linear decline from peak luminosity, similar to that seen in SNe IIL. While we have not included IIL observations or models in the gray clouds of Fig.~\ref{fig:corner}, we would expect $r$-enriched events to differ from IILs in several ways. In addition to the different correlation in $R-K$ vs. $s$ space discussed in the previous paragraph, we expect $v_{50}$ to be greater in $r$-enriched events since the photosphere in IIL explosions is not stalled in the faster moving outer ejecta layers. Furthermore, the linear $r$-enriched models are dimmer than SNe IIP over the first $\sim 100$ days while many observed IIL’s tend to be more luminous than IIP’s \citep{Patat+94, Li+11, Valenti+2016}. Spectral signatures of $r$-process elements may also be effective in breaking degeneracies.

Unsurprisingly, these metrics confirm that the detectability prospect of $r$-process-enriched events increases with the level of enrichment. Observations of luminosity and photosphere velocity at around 50 days after explosion may provide the best opportunity to identify candidate events, while respecting preferences for earlier measurements (relative to the timescales of $r$-process enrichment effects) when events are brighter. Follow up observations would then be required to determine the plateau duration, decline rate, and/or color at day 100, which would all aid in further validating or excluding initial candidates. Of course, if a densely sampled light-curve is obtained up to $\approx 150-200$ days, it would be possible to directly compare to light-curve models (Sec.~\ref{sec:paramstudy}) and estimate the yield of $r$-process material. We note that the more readily observable $V$-band luminosities evolve similarly to the bolometric luminosities (Fig.~\ref{fig:grid}). Thus, $V$-band magnitude at day 50 may be substituted for the metric $L_{50}$ to reveal similar trends in the respective parameter spaces of Fig.~\ref{fig:corner}.

\begin{figure*}
    \centering
    \includegraphics[width=0.9\textwidth]{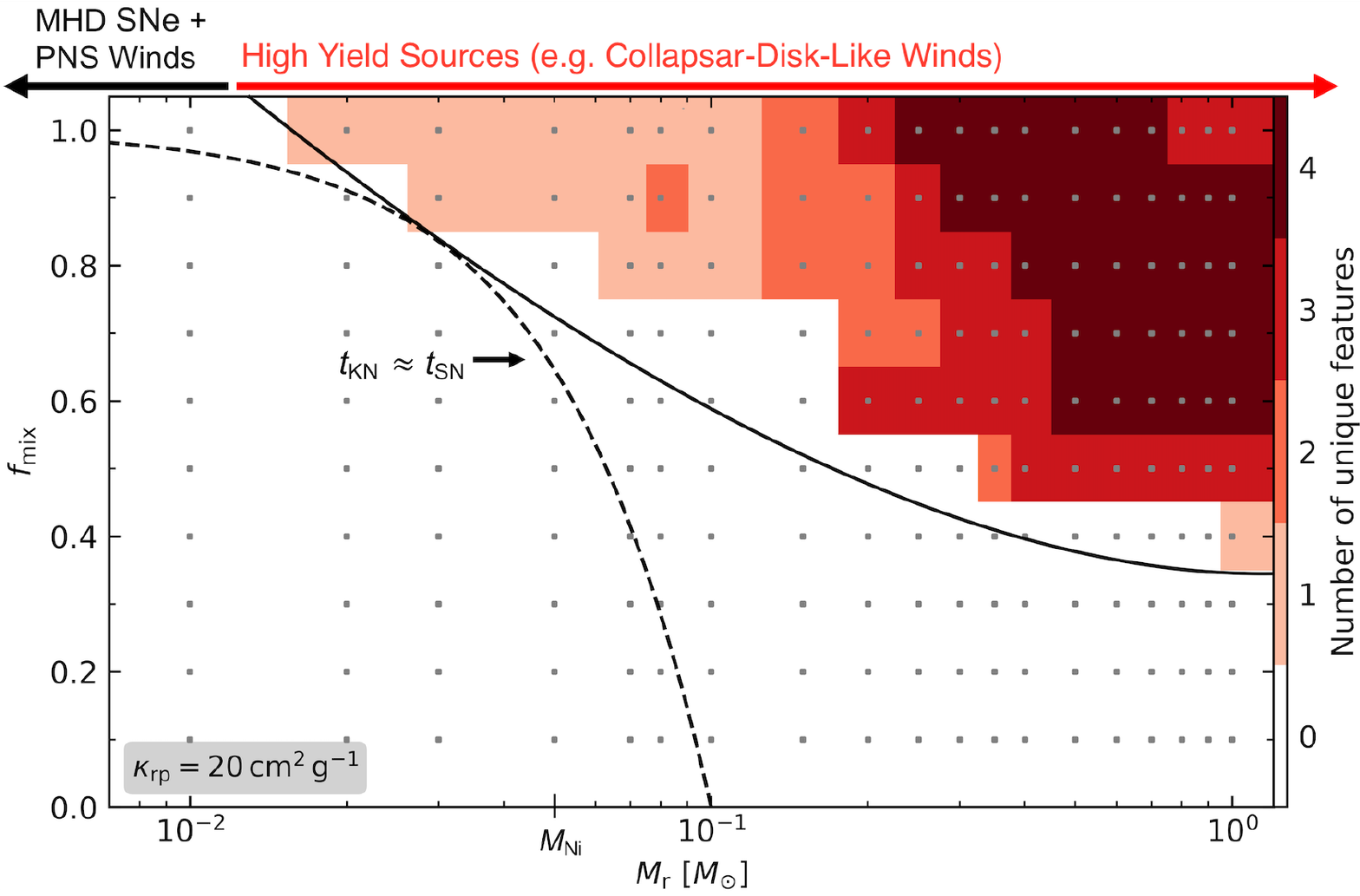}
    \caption{ \normalsize The space of $r$-process mass $M_{\rm r}$ and radial mixing fraction $f_{\rm mix}$, where each grid point (small gray squares) represents an enriched model. Each model is assigned a value between 0 to 4, based on the number of parameter spaces out of the set \{$L_{50} \text{ vs. } t_{\rm p}$, $ v_{50} \text{ vs. } L_{50}$, $s \text{ vs. } t_{\rm p}$, $\text{ and}$, $(R-K)_{100} \text{ vs. } s$\} in which the model is not degenerate with any of the unenriched models (see Fig.~\ref{fig:corner}). A linear-nearest interpolation is performed with these weight values to fill the whole grid. The solid line is drawn qualitatively to encompass the detectable region. The dashed line at which $t_{\rm KN} \approx t_{\rm SN}$ (Eqs.~\eqref{eq: t_kn},~\eqref{eq: t_sn}) is labelled and roughly delineates the detectability boundary for the assumed opacity $\kappa_{\rm rp} = 20 \,\mathrm{cm^2}\,\mathrm{g^{-1}}$. For reference, we also denote the $^{56}$Ni mass of the fiducial model on the $M_{\rm r}$ axis. The approximate ranges of $r$-process mass yields of candidate sources are labelled along the top axis, with PNS winds producing $M_{\rm r} \ll 10^{-3} M_{\odot}$, proto-magnetar winds/MHD SNe producing $M_{\rm r} \lesssim 10^{-2} M_{\odot}$, and a collapsar disk wind-like mechanism producing $M_{\rm r} \gtrsim 10^{-2} M_{\odot}$.}
    \label{fig:detectability}
\end{figure*}

In Fig.~\ref{fig:detectability} we attempt to quantify those regions of $M_{\rm r}-f_{\rm mix}$ parameter space for which $r$-process enrichment produces a significant deviation from an ensemble of unenriched models. Small gray squares represent the $r$-process-enriched models, while the surrounding color denotes the total number of 2D parameter spaces (out of the subset of four spaces bolded in Fig.~\ref{fig:corner}) in which the model deviates significantly from any one of the unenriched models (defined somewhat arbitrarily by the enriched model not overlapping with the grey region). The weight given to each model on the $M_{\rm r} - f_{\rm mix}$ grid (with values ranging from 0 to 4, as in the color bar of Fig.~\ref{fig:detectability}) is interpolated to fill the whole grid. As expected, the upper right hand corner, corresponding to large $M_{\rm r}$ and $f_{\rm mix}$, hosts the most readily distinguishable models. In the next section we describe an approximate analytic estimate for the boundary of this region.

We conclude by noting that, even for models outside this ``detectability'' window, other signatures of $r$-process enrichment may still be observable. Spectroscopic line signatures from $r$-process elements may be present during the early photospheric (e.g., \citealt{Kasen+13}), or late nebular (e.g., \citealt{Hotokezaka+21}), phases, even when photometric signatures are more subtle. Furthermore, events with extremely large but centrally concentrated $r$-process enrichment (high $M_{\rm r}$ but low $f_{\rm mix}$) could power sustained late-time {\it infrared} plateaus (see Sec.~\ref{sec:paramstudy}, \citealt{Siegel+22}) which are not readily apparent from earlier optical light-curve or spectroscopic data.

\subsection{Analytic Estimates} \label{sec:analytic}
The effect of mixing a sufficiently large quantity of $r$-process material\footnote{Such that the opacity of the $r$-process material exceeds that of the non-$r$-enriched material it mixes with. This approximately requires $M_{\rm r}/M_{\rm r,mix} \gtrsim \kappa_{\rm nrp}/\kappa_{\rm rp} \sim 10^{-3}-10^{-2}$ for $\kappa_{\rm nrp} \sim \kappa_{\rm es}$.} out to a given ejecta mass-coordinate $M_{\rm r,mix}$ can in effect be thought of as creating a ``two-layered'' transient: a kilonova (of ejecta mass $\approx M_{\rm r}$ and opacity $\approx \kappa_{\rm rp}$) ``inside" a SN (of ejecta mass $M_{\rm nr} \approx M_{\rm tot}  - M_{\rm r,mix}$ with ordinary electron scattering opacity); see Fig.~\ref{fig:cartoon}.

For the SN portion of the transient, the plateau phase duration can be estimated based on standard analytic arguments \citep{Popov93,Sukhbold+16}
\begin{eqnarray} \label{eq: t_sn}
t_{\rm SN} \simeq 65\,{\rm d}\,\left(\frac{M_{\rm nr}}{5M_{\odot}}\right)^{1/3}\left(\frac{R_0}{500R_{\odot}}\right)^{1/6}\left(\frac{v_{\rm nrp}}{5000\,{\rm km\,s^{-1}}}\right)^{-1/3},
\end{eqnarray}
as determined by the timescale for hydrogen recombination to move inwards in mass coordinate through the ejecta and reduce the opacity (neglecting heat input from the $^{56}$Ni decay chain), where $v_{\rm nrp} \equiv v[(M_{\rm ej} + M_{\rm NS} - M_{\rm r,mix})/2 + M_{\rm r, mix}]$ is the characteristic velocity of the unenriched $r$-process layers. By contrast, during the ``kilonova" phase of the transient, the opacity can be taken as roughly constant, with the escaping emission instead peaking when the expansion and photon diffusion timescales become equal (e.g., \citealt{Arnett80,Metzger+10}),
\begin{eqnarray} \label{eq: t_kn}
&& t_{\rm KN} \approx \left(\frac{\kappa_{\rm rp}M_{\rm r}}{4\pi v_{\rm r} c}\right)^{1/2} \nonumber \\
&\approx& 38\,{\rm d}\,\left(\frac{M_{\rm r}}{0.1M_{\odot}}\right)^{1/2}\left(\frac{\kappa_{\rm rp}}{10\,{\rm cm^{2}\,g^{-1}}}\right)^{1/2}\left(\frac{v_{\rm r}}{5000\,{\rm km\,s^{-1}}}\right)^{-1/2},
\end{eqnarray}
where $v_{\rm r} \equiv v(M_{\rm r}/2 + M_{\rm NS})$ is the characteristic velocity of the slower inner $r$-process-enriched layers.

The above estimates reveal that, for a sufficiently high $r$-process enrichment (i.e., large $M_{\rm r}/M_{\rm nr}$), one can have $t_{\rm KN} \gtrsim t_{\rm SN}$. In such cases, emission from the ``kilonova'' core will peak after the end of the plateau, potentially leading to a double-peaked bolometric light-curve (though likely still single-peaked at optical bands, due to the photosphere temperature being lower during the ``kilonova'' phase than the earlier H-recombination plateau). By contrast, for $t_{\rm KN} \ll t_{\rm SN}$, the kilonova will already have begun to decay by the end of the plateau, its effect on the subsequent light-curve thus being less prominent (see also \citealt{Barnes&Metzger22}). However, the decaying kilonova luminosity may yet exceed the Ni-decay luminosity of the ordinary nebular phase, offering a potential contribution to the late-time excess emission of the $r$-enriched models discussed in Sec.~\ref{sec:results}. We find that the $t_{\rm KN} = t_{\rm SN}$ contour can be approximately represented as $M_{\rm r} \approx 0.1(1-f_{\rm mix})^{2/3}$, as shown with the dashed line in Fig.~\ref{fig:detectability}, where our calculation makes use of velocities, $v_{\rm nrp}$ and $v_{\rm r}$ at $t \approx 100$ days, obtained from \texttt{SNEC} for the fiducial model. The analytic contour is seen to roughly coincide with the ``detectable'' $M_{\rm r}$ boundary (solid line) for well-mixed ejecta ($f_{\rm mix} \gtrsim 0.5$), however, it deviates for lower radial mixing fractions. These low mixing fractions correspond to a kilonova hidden well inside the ejecta core, which would not substantially affect the plateau phase (e.g. Fig.~\ref{fig:grid}). Plateau metrics ($s$, $t_{\rm p}$, $v_{50}$, and $L_{50}$) would thus be unaffected for low $f_{\rm mix}$, explaining why this region appears undetectable despite the analytical prediction. 

\subsection{Implications for the Sites of the $r$-Process}
\label{sec:sites}

Our results indicate that only relatively large $r$-process masses $\gtrsim 10^{-2}-10^{-1}M_{\odot}$ mixed to high mass-coordinates in the SN ejecta produce a discernible effect on the light-curve or other observables (Figs.~\ref{fig:corner}, \ref{fig:detectability}). This minimal detectable $r$-process yield exceeds by $\gtrsim 3$ orders of magnitude those potentially produced in the PNS winds which accompany standard neutrino-driven explosions (e.g., \citealt{Qian&Woosley96}), implying that SN light-curve observations are unlikely to test such models. On the other hand, a small fraction of PNS may be born with extreme properties (e.g., rapidly spinning with ultra-strong magnetic fields $\gtrsim 10^{15}$ G), and the more heavily mass-loaded magneto-centrifugally-driven outflows from such objects can in principle produce substantially higher $r$-process yields $\sim 10^{-2}M_{\odot}$ (e.g., \citealt{Thompson+04,Metzger+07,Winteler+12,Halevi.Mosta_2018.mnras_jet.sn.rproc.3d}). Thus, if $r$-process material mixes outwards into the hydrogen envelope during such ``MHD''-powered explosions, photometric signatures such as those explored here may provide tight constraints for such events. The same is true of collapsar-like events in which accretion disk outflows may generate substantial $r$-process material (e.g., \citealt{Siegel+19}), up to tens of $M_\odot$ in extreme cases \citep{Siegel.ea_2021.arXiv_super.kilonovae}. We label the estimated $r$-process yields of these sources schematically on the top axis of Fig.~\ref{fig:detectability}. 

The source or sources of Galactic $r$-process elements remain a topic of active debate.
While several lines of evidence favor the dominant site being one capable of producing a large per-event $r$-process yield (e.g., \citealt{Wallner+15,Ji+16}), this also implies that $r$-process-enriched SNe must be exceptionally rare. This holds implications for the number of SNe that need be searched to find one enriched event. To explain the entire mass of $r$-process elements (mass number $A \ge 69)$ in the Milky Way via a single formation channel, the product of the per-event $r$-process yield $M_{\rm r}$ and the event rate $\mathcal{R}$ must obey, (e.g., \citealt{Metzger+08a,Hotokezaka+18})
\be
\text{Galactic $r$-process} \implies M_{\rm r} \mathcal{R} \gtrsim 5\times 10^{-7} M_{\odot}\,{\rm yr^{-1}}.
\label{eq:rate}
\ee
Using the present-day Galactic SN rate $\mathcal{R} \approx 0.01-0.02$ yr$^{-1}$ (e.g., \citealt{Rozwadowska+21}) as an estimate for the time-averaged rate, we see that every SN would need only create $M_{\rm r} \sim 10^{-4.5}-10^{-4}M_{\odot}$ to explain all of the Galactic $r$-process if every SN synthesized $r$-process. If instead only a rare subset corresponding to a small fraction $f_{\rm r} \le 1$ of core-collapse SNe generate $r$-process elements (e.g., \citealt{Wallner+15,Ji+16}), and if such a rare SN channel contributes only a fraction $f_{\rm SN} < 1$ of the total $r$-process (the remainder, e.g., arising from neutron star mergers or other sources), then we can modify Eq.~\eqref{eq:rate} to calculate the required yield from each such rare SNe:
\be
M_{\rm r} \sim 10^{-2}M_{\odot}\left(\frac{f_{\rm r}}{10^{-3}}\right)^{-1}\left(\frac{f_{\rm SN}}{0.3}\right).
\label{eq:Mrdetect}
\ee
While Figs.~\ref{fig:corner}, \ref{fig:detectability} show that $r$-process yields $M_{\rm r} \gtrsim 10^{-2}-10^{-1}M_{\odot}$, mixed sufficiently close to the ejecta surface (for $f_{\rm mix} \gtrsim 0.4$), may be sufficient for identification based on the photometry, Eq.~\eqref{eq:Mrdetect} show that such high-yields are necessarily extremely rare among the core-collapse population, to avoid surpassing the Galactic $r$-process abundances. This has the practical consequence that at least $N > 1/f_{\rm r} \sim 10^{3}-10^{4}$ core-collapse SNe would need to be searched to identify a single $r$-process-enriched event (for $M_{\rm r} \gtrsim 10^{-2}-10^{-1} M_{\odot}$), even assuming such events generate a significant fraction (e.g., $f_{\rm SN} \gtrsim 0.3$) of the total Galactic $r$-process budget.

The {\it Rubin Observatory} is expected to discover $N_{\rm Rubin} \sim 10^{6}$ core-collapse SNe over its 10 year survey, potentially yielding a sample of up to $N_{\rm Rubin}f_{\rm r} \sim 10^{2}-10^{3}$ enriched candidates. Once candidates have been identified, confirmation will likely require spectroscopic follow-up observations to search, e.g. for $r$-process line signatures \citep{Kasen+13, Pian+2017}.  Alternatively, the absence of a single $r$-enriched candidate SN in {\it Rubin} would imply a constraint $f_{\rm SN} \lesssim 10^{-2}$ on enrichment events with $M_{\rm r} \sim 0.1M_{\odot}$.

\section{Conclusions}
\label{sec:conclusions}

We have performed an explorative study of the effects of $r$-process enrichment on the light-curves of hydrogen-rich SNe resulting from the core-collapse of single solar-metallicity stars. Using a one-dimensional radiation-hydrodynamics model, we include the effects of the high optical/UV opacity of $r$-process elements (modeled as a constant gray opacity) and their associated radioactive heating rate (Sec.~\ref{sec:model}), the latter augmenting that of the $^{56}$Ni produced in the explosion. The total mass of $r$-process elements, $M_{\rm r}$, and its radial distribution within the ejecta (encapsulated mainly in the radial mixing coordinate $f_{\rm mix}$), are taken as free parameters. We produce a grid of light-curve models spanning a physically motivated range of $M_{\rm r}$ and $f_{\rm mix}$ (Fig.~\ref{fig:grid}), and compare our results to a sample of ordinary (i.e., non-$r$-process-enriched) SNe IIP models and observational data (Fig.~\ref{fig:corner}). 

For an assumed $r$-process opacity $\kappa_{\rm rp} \approx 20$ cm$^{2}\,$g$^{-1}$, we find a lower limit $M_{\rm r} \gtrsim 10^{-2} M_{\odot}$ on the $r$-process yield required to produce an an appreciable change to the bolometric and $V$-band light-curves, and the $R-K$ color evolution, relative to an equivalent unenriched model. Although the value of $\kappa_{\rm rp}$ is uncertain theoretically, we note that it enters our model exclusively through the combination $\kappa_{\rm rp}X_{\rm r}$ where $X_{\rm r} \propto M_{\rm r}$ (Eqs.~\ref{eq:kappar},~\ref{eq:Xr}). This means that our results for models with varying $M_{\rm r}$ can be readily translated to scenarios where $\kappa_{\rm rp}$ is larger or smaller than the assumed value.

The key photometric features of $r$-process enrichment include a reduction in length of the characteristic plateau phase, a corresponding steep transition to the nebular phase, and a late time infrared excess resulting in the reddening of the SN colors (Fig.~\ref{fig:cartoon}, Sec.~\ref{sec:fiducial}). These effects become more pronounced with increasing levels of enrichment. From these signatures, we have identified key metrics that may distinguish $r$-process-enriched events from otherwise ordinary SNe in statistically large samples of SNe observations:
\begin{itemize}
    \item Luminosity and photosphere velocity at day 50 present the best opportunity to identify enriched events relatively early in their evolution.
    \item Measurements of the plateau duration, plateau decline rate, and/or $R - K$ at day 100 can be used to further constrain the viability of initial $r$-process-enriched SN candidates (Sec.~\ref{sec:detectability}).
    \item  Densely sampled light-curves of initial candidates may also confirm $r$-process enrichment via comparison to the models presented in this paper.
\end{itemize}

Based on the considered metrics, we quantify an approximate detectability threshold for $r$-process enrichment (Sec.~\ref{sec:analytic}). This analysis also suggests that events with $M_{\rm r} \gtrsim 10^{-2} M_{\odot}$ may be distinguishable from ordinary SNe IIP of typical progenitor and explosion properties, depending on the ejecta radial mixing fraction. Such detectable $r$-process masses are well above the yields of standard (weakly magnetized) PNS wind models, but may be consistent with rare high-yield events such as MHD-SNe, proto-magnetar winds, or accretion disk outflows in cases of substantial fall-back. Most of these scenarios require high pre-collapse angular momentum in the core of the progenitor star, which in light of efficient angular momentum transport during earlier phases of stellar evolution (e.g., \citealt{Spruit1999, spruit:02, Stello+16,Eggenberger+2019, Fuller+19}) may require a well-timed merger or tidal interaction with a tight binary companion (e.g., \citealt{Cantiello+07,Fuller&Lu22}), which may preclude the presence of a H-envelope at collapse (however, see \citealt{Dopita1988, Chatzopoulos+2020, Wheeler&Chatzaopoulos23,Burrows+23}).

To improve our study, multi-group radiation-hydrodynamics calculations,  which account for more realistic wavelength- and temperature-dependent $r$-process opacities (e.g., within the expansion opacity formalism), would enable more accurate light-curve models. For example, the strong wavelength dependence of the bound-bound opacities of the lanthanide group elements could potentially reveal a favorable increase in infrared range excess (e.g., an increase in $R-K$ values) when multi-group radiative transfer is implemented. In addition, multidimensional (magneto-)hydrodynamic simulations of the interaction between the $r$-process-enriched wind or jet from the central compact object, and the expanding SN ejecta, could better inform our assumptions about the radial and angular distribution of the $r$-process material within the ejecta (e.g., \citealt{Barnes&Duffell23}). Such simulations can be used to constrain the $f_{\rm mix}$ parameter to $M_{\rm r}$ and other properties of the stellar explosion, and eventually inform the (likely non-spherical) distribution of $r$-process enrichment within the ejecta to motivate multi-dimensional transport simulations.

The notable features of $r$-process enrichment predicted in this work may be compared to upcoming observations from \textit{The Rubin Observatory} and \textit{Roman Telescope}. \textit{Rubin's LSST} will present an opportunity to probe statistically large data samples for such rare events and \textit{Roman} will be well suited to search for $r$-process signatures due its sensitivity at near-infrared wavelengths.

\section{Data Availability}
The modified version of $\texttt{SNEC}$, post-processing and visualization tools, and data used in this work are available online at \url{https://zenodo.org/records/10515953}.

\begin{acknowledgements}

We thank Annastasia Haynie, Viktoriya Morozova and Tony Piro for valuable discussions about the \texttt{SNEC} software instrument. We thank Jeniveve Pearson and the anonymous referee for constructive feedback. AP and BDM are supported in part by the National Science Foundation (grant numbers AST-2002577, AST-2009255) and the NASA Fermi Guest Investigator Program (grant number 80NSSC22K1574). The Flatiron Institute is supported by the Simons Foundation.

\end{acknowledgements}

\bibliographystyle{aasjournal}
\bibliography{main_new}

\newcommand{\noop}[1]{}
\begin{thebibliography}{}
\expandafter\ifx\csname natexlab\endcsname\relax\def\natexlab#1{#1}\fi
\providecommand{\url}[1]{\href{#1}{#1}}
\providecommand{\dodoi}[1]{doi:~\href{http://doi.org/#1}{\nolinkurl{#1}}}
\providecommand{\doeprint}[1]{\href{http://ascl.net/#1}{\nolinkurl{http://ascl.net/#1}}}
\providecommand{\doarXiv}[1]{\href{https://arxiv.org/abs/#1}{\nolinkurl{https://arxiv.org/abs/#1}}}

\bibitem[{{Anand} {et~al.}(2023){Anand}, {Barnes}, {Yang}, {Kasliwal}, {Coughlin}, {et~al.}}]{Anand+23}
{Anand}, S., {Barnes}, J., {Yang}, S., {et~al.} 2023, arXiv e-prints, arXiv:2302.09226, \dodoi{10.48550/arXiv.2302.09226}

\bibitem[{{Anderson} {et~al.}(2014){Anderson}, {Gonz{\'a}lez-Gait{\'a}n}, {Hamuy}, {Guti{\'e}rrez}, {Stritzinger}, {Olivares E.}, {Phillips}, {Schulze}, {Antezana}, {Bolt}, {Campillay}, {Castell{\'o}n}, {Contreras}, {de Jaeger}, {Folatelli}, {F{\"o}rster}, {Freedman}, {Gonz{\'a}lez}, {Hsiao}, {Krzemi{\'n}ski}, {Krisciunas}, {Maza}, {McCarthy}, {Morrell}, {Persson}, {Roth}, {Salgado}, {Suntzeff}, \& {Thomas-Osip}}]{Anderson+14}
{Anderson}, J.~P., {Gonz{\'a}lez-Gait{\'a}n}, S., {Hamuy}, M., {et~al.} 2014, \apj, 786, 67, \dodoi{10.1088/0004-637X/786/1/67}

\bibitem[{{Arnett}(1980)}]{Arnett80}
{Arnett}, W.~D. 1980, \apj, 237, 541, \dodoi{10.1086/157898}

\bibitem[{Baldeschi {et~al.}(2020)Baldeschi, Miller, Stroh, Margutti, \& Coppejans}]{Baldeschi+2020}
Baldeschi, A., Miller, A., Stroh, M., Margutti, R., \& Coppejans, D.~L. 2020, The Astrophysical Journal, 902, 60, \dodoi{10.3847/1538-4357/abb1c0}

\bibitem[{{Barker} {et~al.}(2023){Barker}, {O'Connor}, \& {Couch}}]{Barker+23}
{Barker}, B.~L., {O'Connor}, E.~P., \& {Couch}, S.~M. 2023, \apjl, 944, L2, \dodoi{10.3847/2041-8213/acb052}

\bibitem[{{Barnes} \& {Duffell}(2023)}]{Barnes&Duffell23}
{Barnes}, J., \& {Duffell}, P.~C. 2023, arXiv e-prints, arXiv:2305.00056, \dodoi{10.48550/arXiv.2305.00056}

\bibitem[{{Barnes} \& {Kasen}(2013)}]{Barnes_2013}
{Barnes}, J., \& {Kasen}, D. 2013, \apj, 775, 18, \dodoi{10.1088/0004-637X/775/1/18}

\bibitem[{{Barnes} {et~al.}(2016){Barnes}, {Kasen}, {Wu}, \& {Mart{\'{\i}}nez-Pinedo}}]{Barnes_etal_2016}
{Barnes}, J., {Kasen}, D., {Wu}, M.-R., \& {Mart{\'{\i}}nez-Pinedo}, G. 2016, \apj, 829, 110, \dodoi{10.3847/0004-637X/829/2/110}

\bibitem[{{Barnes} \& {Metzger}(2022)}]{Barnes&Metzger22}
{Barnes}, J., \& {Metzger}, B.~D. 2022, \apjl, 939, L29, \dodoi{10.3847/2041-8213/ac9b41}

\bibitem[{Barnes \& Metzger(2023)}]{Barnes&Metzger23}
Barnes, J., \& Metzger, B.~D. 2023, The Astrophysical Journal, 947, 55, \dodoi{10.3847/1538-4357/acc384}

\bibitem[{{Bellm} {et~al.}(2019){Bellm}, {Kulkarni}, {Graham}, {et~al.}}]{Bellm+19}
{Bellm}, E.~C., {Kulkarni}, S.~R., {Graham}, M.~J., {et~al.} 2019, \pasp, 131, 018002, \dodoi{10.1088/1538-3873/aaecbe}

\bibitem[{{Bersten}(2013)}]{Bersten2013}
{Bersten}, M.~C. 2013, arXiv e-prints, arXiv:1303.0639, \dodoi{10.48550/arXiv.1303.0639}

\bibitem[{{Bersten} {et~al.}(2011){Bersten}, {Benvenuto}, \& {Hamuy}}]{Bersten2011}
{Bersten}, M.~C., {Benvenuto}, O., \& {Hamuy}, M. 2011, \apj, 729, 61, \dodoi{10.1088/0004-637X/729/1/61}

\bibitem[{{Blinnikov} \& {Bartunov}(1993)}]{Blinnikov+Bartunov93}
{Blinnikov}, S.~I., \& {Bartunov}, O.~S. 1993, \aap, 273, 106

\bibitem[{{Blinnikov} {et~al.}(1998){Blinnikov}, {Eastman}, {Bartunov}, {Popolitov}, \& {Woosley}}]{Blinnikov+98}
{Blinnikov}, S.~I., {Eastman}, R., {Bartunov}, O.~S., {Popolitov}, V.~A., \& {Woosley}, S.~E. 1998, \apj, 496, 454, \dodoi{10.1086/305375}

\bibitem[{{Brown} {et~al.}(2013){Brown}, {Baliber}, {Bianco}, {Bowman}, {Burleson}, {Conway}, {Crellin}, {Depagne}, {De Vera}, {Dilday}, {Dragomir}, {Dubberley}, {Eastman}, {Elphick}, {Falarski}, {Foale}, {Ford}, {Fulton}, {Garza}, {Gomez}, {Graham}, {Greene}, {Haldeman}, {Hawkins}, {Haworth}, {Haynes}, {Hidas}, {Hjelstrom}, {Howell}, {Hygelund}, {Lister}, {Lobdill}, {Martinez}, {Mullins}, {Norbury}, {Parrent}, {Paulson}, {Petry}, {Pickles}, {Posner}, {Rosing}, {Ross}, {Sand}, {Saunders}, {Shobbrook}, {Shporer}, {Street}, {Thomas}, {Tsapras}, {Tufts}, {Valenti}, {Vander Horst}, {Walker}, {White}, \& {Willis}}]{Brown2013}
{Brown}, T.~M., {Baliber}, N., {Bianco}, F.~B., {et~al.} 2013, \pasp, 125, 1031, \dodoi{10.1086/673168}

\bibitem[{{Burrows} {et~al.}(1995){Burrows}, {Hayes}, \& {Fryxell}}]{Burrows+95}
{Burrows}, A., {Hayes}, J., \& {Fryxell}, B.~A. 1995, \apj, 450, 830, \dodoi{10.1086/176188}

\bibitem[{{Burrows} {et~al.}(2023){Burrows}, {Vartanyan}, \& {Wang}}]{Burrows+23}
{Burrows}, A., {Vartanyan}, D., \& {Wang}, T. 2023, \apj, 957, 68, \dodoi{10.3847/1538-4357/acfc1c}

\bibitem[{{Cantiello} {et~al.}(2007){Cantiello}, {Yoon}, {Langer}, \& {Livio}}]{Cantiello+07}
{Cantiello}, M., {Yoon}, S.~C., {Langer}, N., \& {Livio}, M. 2007, \aap, 465, L29, \dodoi{10.1051/0004-6361:20077115}

\bibitem[{Chatzopoulos {et~al.}(2020)Chatzopoulos, Frank, Marcello, \& Clayton}]{Chatzopoulos+2020}
Chatzopoulos, E., Frank, J., Marcello, D.~C., \& Clayton, G.~C. 2020, The Astrophysical Journal, 896, 50, \dodoi{10.3847/1538-4357/ab91bb}

\bibitem[{{C{\^o}t{\'e}} {et~al.}(2019){C{\^o}t{\'e}}, {Eichler}, {Arcones}, {Hansen}, {Simonetti}, {Frebel}, {Fryer}, {Pignatari}, {Reichert}, {Belczynski}, \& {Matteucci}}]{Cote+19}
{C{\^o}t{\'e}}, B., {Eichler}, M., {Arcones}, A., {et~al.} 2019, \apj, 875, 106, \dodoi{10.3847/1538-4357/ab10db}

\bibitem[{{Cowan} {et~al.}(2021){Cowan}, {Sneden}, {Lawler}, {Aprahamian}, {Wiescher}, {Langanke}, {Mart{\'\i}nez-Pinedo}, \& {Thielemann}}]{Cowan+21}
{Cowan}, J.~J., {Sneden}, C., {Lawler}, J.~E., {et~al.} 2021, Reviews of Modern Physics, 93, 015002, \dodoi{10.1103/RevModPhys.93.015002}

\bibitem[{{Crowther}(2007)}]{Crowther2007}
{Crowther}, P.~A. 2007, \araa, 45, 177, \dodoi{10.1146/annurev.astro.45.051806.110615}

\bibitem[{{de Jager} {et~al.}(1988){de Jager}, {Nieuwenhuijzen}, \& {van der Hucht}}]{deJager+1988}
{de Jager}, C., {Nieuwenhuijzen}, H., \& {van der Hucht}, K.~A. 1988, \aaps, 72, 259

\bibitem[{{Desai} {et~al.}(2022){Desai}, {Siegel}, \& {Metzger}}]{Desai+22}
{Desai}, D., {Siegel}, D.~M., \& {Metzger}, B.~D. 2022, \apj, 931, 104, \dodoi{10.3847/1538-4357/ac69da}

\bibitem[{{Desai} {et~al.}(2023){Desai}, {Siegel}, \& {Metzger}}]{Desai+23}
{Desai}, D.~K., {Siegel}, D.~M., \& {Metzger}, B.~D. 2023, \apj, 954, 192, \dodoi{10.3847/1538-4357/acea83}

\bibitem[{{Dessart} \& {Audit}(2018)}]{Dessart+18}
{Dessart}, L., \& {Audit}, E. 2018, \aap, 613, A5, \dodoi{10.1051/0004-6361/201732229}

\bibitem[{{Dessart} \& {Hillier}(2019)}]{Dessart+19}
{Dessart}, L., \& {Hillier}, D.~J. 2019, \aap, 625, A9, \dodoi{10.1051/0004-6361/201834732}

\bibitem[{Dopita(1988)}]{Dopita1988}
Dopita, M.~A. 1988, Nature, 331, 506, \dodoi{10.1038/331506a0}

\bibitem[{{Duffell}(2016)}]{Duffell2016}
{Duffell}, P.~C. 2016, \apj, 821, 76, \dodoi{10.3847/0004-637X/821/2/76}

\bibitem[{{Eastman} {et~al.}(1994){Eastman}, {Woosley}, {Weaver}, \& {Pinto}}]{Eastman1994}
{Eastman}, R.~G., {Woosley}, S.~E., {Weaver}, T.~A., \& {Pinto}, P.~A. 1994, \apj, 430, 300, \dodoi{10.1086/174404}

\bibitem[{{Eggenberger} {et~al.}(2019){Eggenberger}, {Buldgen}, \& {Salmon}}]{Eggenberger+2019}
{Eggenberger}, P., {Buldgen}, G., \& {Salmon}, S.~J.~A.~J. 2019, \aap, 626, L1, \dodoi{10.1051/0004-6361/201935509}

\bibitem[{{Falk} \& {Arnett}(1977)}]{Falk&Arnett1977}
{Falk}, S.~W., \& {Arnett}, W.~D. 1977, \apjs, 33, 515, \dodoi{10.1086/190440}

\bibitem[{{Fassia} \& {Meikle}(1999)}]{Fassia+1999}
{Fassia}, A., \& {Meikle}, W.~P.~S. 1999, \mnras, 302, 314, \dodoi{10.1046/j.1365-8711.1999.02127.x}

\bibitem[{{Ferguson} {et~al.}(2005){Ferguson}, {Alexander}, {Allard}, {Barman}, {Bodnarik}, {Hauschildt}, {Heffner-Wong}, \& {Tamanai}}]{Ferguson2005}
{Ferguson}, J.~W., {Alexander}, D.~R., {Allard}, F., {et~al.} 2005, \apj, 623, 585, \dodoi{10.1086/428642}

\bibitem[{Foley \& Mandel(2013)}]{Foley+Mandel2013}
Foley, R.~J., \& Mandel, K. 2013, The Astrophysical Journal, 778, 167, \dodoi{10.1088/0004-637X/778/2/167}

\bibitem[{{Fraser}(2020)}]{Fraser20}
{Fraser}, M. 2020, Royal Society Open Science, 7, 200467, \dodoi{10.1098/rsos.200467}

\bibitem[{{Fujibayashi} {et~al.}(2020){Fujibayashi}, {Shibata}, {Wanajo}, {Kiuchi}, {Kyutoku}, \& {Sekiguchi}}]{Fujibayashi+20}
{Fujibayashi}, S., {Shibata}, M., {Wanajo}, S., {et~al.} 2020, \prd, 102, 123014, \dodoi{10.1103/PhysRevD.102.123014}

\bibitem[{{Fuller} \& {Lu}(2022)}]{Fuller&Lu22}
{Fuller}, J., \& {Lu}, W. 2022, \mnras, 511, 3951, \dodoi{10.1093/mnras/stac317}

\bibitem[{{Fuller} {et~al.}(2019){Fuller}, {Piro}, \& {Jermyn}}]{Fuller+19}
{Fuller}, J., {Piro}, A.~L., \& {Jermyn}, A.~S. 2019, \mnras, 485, 3661, \dodoi{10.1093/mnras/stz514}

\bibitem[{{Gagliano} {et~al.}(2023){Gagliano}, {Contardo}, {Foreman-Mackey}, {Malz}, \& {Aleo}}]{Gagliano+23}
{Gagliano}, A., {Contardo}, G., {Foreman-Mackey}, D., {Malz}, A.~I., \& {Aleo}, P.~D. 2023, \apj, 954, 6, \dodoi{10.3847/1538-4357/ace326}

\bibitem[{{Goldberg} \& {Bildsten}(2020)}]{GB2020}
{Goldberg}, J.~A., \& {Bildsten}, L. 2020, \apjl, 895, L45, \dodoi{10.3847/2041-8213/ab9300}

\bibitem[{{Goldberg} {et~al.}(2019){Goldberg}, {Bildsten}, \& {Paxton}}]{Goldberg+19}
{Goldberg}, J.~A., {Bildsten}, L., \& {Paxton}, B. 2019, \apj, 879, 3, \dodoi{10.3847/1538-4357/ab22b6}

\bibitem[{{Grassberg} {et~al.}(1971){Grassberg}, {Imshennik}, \& {Nadyozhin}}]{Grassberg+1971}
{Grassberg}, E.~K., {Imshennik}, V.~S., \& {Nadyozhin}, D.~K. 1971, \apss, 10, 28, \dodoi{10.1007/BF00654604}

\bibitem[{Gutierrez {et~al.}(2017)Gutierrez, Anderson, Hamuy, González-Gaitan, Galbany, Dessart, Stritzinger, Phillips, Morrell, \& Folatelli}]{Gutierrez+17}
Gutierrez, C.~P., Anderson, J.~P., Hamuy, M., {et~al.} 2017, The Astrophysical Journal, 850, 90, \dodoi{10.3847/1538-4357/aa8f42}

\bibitem[{{Halevi} \& {M{\"o}sta}(2018)}]{Halevi.Mosta_2018.mnras_jet.sn.rproc.3d}
{Halevi}, G., \& {M{\"o}sta}, P. 2018, \mnras, 477, 2366, \dodoi{10.1093/mnras/sty797}

\bibitem[{Hamuy(2003)}]{Hamuy2003}
Hamuy, M. 2003, The Astrophysical Journal, 582, 905, \dodoi{10.1086/344689}

\bibitem[{{Herant} \& {Woosley}(1994)}]{Herant1994}
{Herant}, M., \& {Woosley}, S.~E. 1994, \apj, 425, 814, \dodoi{10.1086/174026}

\bibitem[{{Hiramatsu} {et~al.}(2021){Hiramatsu}, {Howell}, {Moriya}, {Goldberg}, {Hosseinzadeh}, {Arcavi}, {Anderson}, {Guti{\'e}rrez}, {Burke}, {McCully}, {Valenti}, {Galbany}, {Fang}, {Maeda}, {Folatelli}, {Hsiao}, {Morrell}, {Phillips}, {Stritzinger}, {Suntzeff}, {Gromadzki}, {Maguire}, {M{\"u}ller-Bravo}, \& {Young}}]{Hiramatsu2021b}
{Hiramatsu}, D., {Howell}, D.~A., {Moriya}, T.~J., {et~al.} 2021, \apj, 913, 55, \dodoi{10.3847/1538-4357/abf6d6}

\bibitem[{{Horowitz} {et~al.}(2019){Horowitz}, {Arcones}, {C{\^o}t{\'e}}, {Dillmann}, {Nazarewicz}, {Roederer}, {Schatz}, {Aprahamian}, {Atanasov}, {Bauswein}, {Beers}, {Bliss}, {Brodeur}, {Clark}, {Frebel}, {Foucart}, {Hansen}, {Just}, {Kankainen}, {McLaughlin}, {Kelly}, {Liddick}, {Lee}, {Lippuner}, {Martin}, {Mendoza-Temis}, {Metzger}, {Mumpower}, {Perdikakis}, {Pereira}, {O'Shea}, {Reifarth}, {Rogers}, {Siegel}, {Spyrou}, {Surman}, {Tang}, {Uesaka}, \& {Wang}}]{Horowitz+19}
{Horowitz}, C.~J., {Arcones}, A., {C{\^o}t{\'e}}, B., {et~al.} 2019, Journal of Physics G Nuclear Physics, 46, 083001, \dodoi{10.1088/1361-6471/ab0849}

\bibitem[{{Hotokezaka} {et~al.}(2018){Hotokezaka}, {Beniamini}, \& {Piran}}]{Hotokezaka+18}
{Hotokezaka}, K., {Beniamini}, P., \& {Piran}, T. 2018, International Journal of Modern Physics D, 27, 1842005, \dodoi{10.1142/S0218271818420051}

\bibitem[{{Hotokezaka} {et~al.}(2021){Hotokezaka}, {Tanaka}, {Kato}, \& {Gaigalas}}]{Hotokezaka+21}
{Hotokezaka}, K., {Tanaka}, M., {Kato}, D., \& {Gaigalas}, G. 2021, arXiv e-prints, arXiv:2102.07879.
\newblock \doarXiv{2102.07879}

\bibitem[{{Iglesias} \& {Rogers}(1996)}]{Iglesias&Rogers1996}
{Iglesias}, C.~A., \& {Rogers}, F.~J. 1996, \apj, 464, 943, \dodoi{10.1086/177381}

\bibitem[{{Ivezi{\'c}} {et~al.}(2019){Ivezi{\'c}}, {Kahn}, {Tyson}, {et~al.}}]{Ivezic+19}
{Ivezi{\'c}}, {\v{Z}}., {Kahn}, S.~M., {Tyson}, J.~A., {et~al.} 2019, \apj, 873, 111, \dodoi{10.3847/1538-4357/ab042c}

\bibitem[{{Ji} {et~al.}(2016){Ji}, {Frebel}, {Chiti}, \& {Simon}}]{Ji+16}
{Ji}, A.~P., {Frebel}, A., {Chiti}, A., \& {Simon}, J.~D. 2016, Nature, 531, 610, \dodoi{10.1038/nature17425}

\bibitem[{{Just} {et~al.}(2022){Just}, {Aloy}, {Obergaulinger}, \& {Nagataki}}]{Just+22}
{Just}, O., {Aloy}, M.~A., {Obergaulinger}, M., \& {Nagataki}, S. 2022, \apjl, 934, L30, \dodoi{10.3847/2041-8213/ac83a1}

\bibitem[{{Kasen} {et~al.}(2013){Kasen}, {Badnell}, \& {Barnes}}]{Kasen+13}
{Kasen}, D., {Badnell}, N.~R., \& {Barnes}, J. 2013, ApJ, submitted, arXiv:1303.5788.
\newblock \doarXiv{1303.5788}

\bibitem[{{Kasen} \& {Bildsten}(2010)}]{Kasen&Bildsten10}
{Kasen}, D., \& {Bildsten}, L. 2010, \apj, 717, 245, \dodoi{10.1088/0004-637X/717/1/245}

\bibitem[{{Kasen} {et~al.}(2009){Kasen}, {R{\"o}pke}, \& {Woosley}}]{Kasen_2009}
{Kasen}, D., {R{\"o}pke}, F.~K., \& {Woosley}, S.~E. 2009, \nat, 460, 869, \dodoi{10.1038/nature08256}

\bibitem[{{Kasen} \& {Woosley}(2009)}]{Kasen&Woosley2009}
{Kasen}, D., \& {Woosley}, S.~E. 2009, \apj, 703, 2205, \dodoi{10.1088/0004-637X/703/2/2205}

\bibitem[{{Kochanek} {et~al.}(2017){Kochanek}, {Shappee}, {Stanek}, {Holoien}, {Thompson}, {Prieto}, {Dong}, {Shields}, {Will}, {Britt}, {Perzanowski}, \& {Pojma{\'n}ski}}]{Kochanek+17}
{Kochanek}, C.~S., {Shappee}, B.~J., {Stanek}, K.~Z., {et~al.} 2017, \pasp, 129, 104502, \dodoi{10.1088/1538-3873/aa80d9}

\bibitem[{Li {et~al.}(2011)Li, Leaman, Chornock, Filippenko, Poznanski, Ganeshalingam, Wang, Modjaz, Jha, Foley, \& Smith}]{Li+11}
Li, W., Leaman, J., Chornock, R., {et~al.} 2011, Monthly Notices of the Royal Astronomical Society, 412, 1441, \dodoi{10.1111/j.1365-2966.2011.18160.x}

\bibitem[{{Lin} {et~al.}(2023){Lin}, {Zhang}, \& {Zhang}}]{Lin+23}
{Lin}, H., {Zhang}, J., \& {Zhang}, X. 2023, Universe, 9, 201, \dodoi{10.3390/universe9050201}

\bibitem[{{Litvinova} \& {Nadezhin}(1985)}]{Litvinova&Nadezhin1985}
{Litvinova}, I.~Y., \& {Nadezhin}, D.~K. 1985, Soviet Astronomy Letters, 11, 145

\bibitem[{{LSST Science Collaboration} {et~al.}(2009)}]{LSST+09}
{LSST Science Collaboration}, {et~al.} 2009, arXiv e-prints, arXiv:0912.0201, \dodoi{10.48550/arXiv.0912.0201}

\bibitem[{{Ma} \& {Fuller}(2019)}]{Ma+Fuller2019}
{Ma}, L., \& {Fuller}, J. 2019, \mnras, 488, 4338, \dodoi{10.1093/mnras/stz2009}

\bibitem[{{MacFadyen} \& {Woosley}(1999)}]{MacFadyen&Woosley99}
{MacFadyen}, A.~I., \& {Woosley}, S.~E. 1999, \apj, 524, 262, \dodoi{10.1086/307790}

\bibitem[{{Metzger} {et~al.}(2018){Metzger}, {Beniamini}, \& {Giannios}}]{Metzger+18b}
{Metzger}, B.~D., {Beniamini}, P., \& {Giannios}, D. 2018, \apj, 857, 95, \dodoi{10.3847/1538-4357/aab70c}

\bibitem[{{Metzger} {et~al.}(2008){Metzger}, {Quataert}, \& {Thompson}}]{Metzger+08a}
{Metzger}, B.~D., {Quataert}, E., \& {Thompson}, T.~A. 2008, \mnras, 385, 1455, \dodoi{10.1111/j.1365-2966.2008.12923.x}

\bibitem[{{Metzger} {et~al.}(2007){Metzger}, {Thompson}, \& {Quataert}}]{Metzger+07}
{Metzger}, B.~D., {Thompson}, T.~A., \& {Quataert}, E. 2007, \apj, 659, 561, \dodoi{10.1086/512059}

\bibitem[{{Metzger} {et~al.}(2010){Metzger}, {Mart{\'{\i}}nez-Pinedo}, {Darbha}, {Quataert}, {Arcones}, {Kasen}, {Thomas}, {Nugent}, {Panov}, \& {Zinner}}]{Metzger+10}
{Metzger}, B.~D., {Mart{\'{\i}}nez-Pinedo}, G., {Darbha}, S., {et~al.} 2010, Mon. Not. R. Astron. Soc., 406, 2650, \dodoi{10.1111/j.1365-2966.2010.16864.x}

\bibitem[{{Meyer} {et~al.}(1992){Meyer}, {Mathews}, {Howard}, {Woosley}, \& {Hoffman}}]{Meyer+92}
{Meyer}, B.~S., {Mathews}, G.~J., {Howard}, W.~M., {Woosley}, S.~E., \& {Hoffman}, R.~D. 1992, \apj, 399, 656, \dodoi{10.1086/171957}

\bibitem[{{Miller} {et~al.}(2020){Miller}, {Sprouse}, {Fryer}, {Ryan}, {Dolence}, {Mumpower}, \& {Surman}}]{Miller+20}
{Miller}, J.~M., {Sprouse}, T.~M., {Fryer}, C.~L., {et~al.} 2020, \apj, 902, 66, \dodoi{10.3847/1538-4357/abb4e3}

\bibitem[{Moriya {et~al.}(2016)Moriya, Pruzhinskaya, Ergon, \& Blinnikov}]{Moriya+2016}
Moriya, T.~J., Pruzhinskaya, M.~V., Ergon, M., \& Blinnikov, S.~I. 2016, Monthly Notices of the Royal Astronomical Society, 455, 423, \dodoi{10.1093/mnras/stv2336}

\bibitem[{{Morozova} {et~al.}(2016){Morozova}, {Piro}, {Renzo}, \& {Ott}}]{Morozova+2016}
{Morozova}, V., {Piro}, A.~L., {Renzo}, M., \& {Ott}, C.~D. 2016, \apj, 829, 109, \dodoi{10.3847/0004-637X/829/2/109}

\bibitem[{{Morozova} {et~al.}(2015){Morozova}, {Piro}, {Renzo}, {Ott}, {Clausen}, {Couch}, {Ellis}, \& {Roberts}}]{Morozova15}
{Morozova}, V., {Piro}, A.~L., {Renzo}, M., {et~al.} 2015, \apj, 814, 63, \dodoi{10.1088/0004-637X/814/1/63}

\bibitem[{Morozova {et~al.}(2017)Morozova, Piro, \& Valenti}]{Morozova_2017}
Morozova, V., Piro, A.~L., \& Valenti, S. 2017, The Astrophysical Journal, 838, 28, \dodoi{10.3847/1538-4357/aa6251}

\bibitem[{{M{\"o}sta} {et~al.}(2014){M{\"o}sta}, {Richers}, {Ott}, {Haas}, {Piro}, {Boydstun}, {Abdikamalov}, {Reisswig}, \& {Schnetter}}]{Mosta+14}
{M{\"o}sta}, P., {Richers}, S., {Ott}, C.~D., {et~al.} 2014, \apjl, 785, L29, \dodoi{10.1088/2041-8205/785/2/L29}

\bibitem[{{M{\"u}ller} {et~al.}(2017){M{\"u}ller}, {Prieto}, {Pejcha}, \& {Clocchiatti}}]{Muller2017}
{M{\"u}ller}, T., {Prieto}, J.~L., {Pejcha}, O., \& {Clocchiatti}, A. 2017, \apj, 841, 127, \dodoi{10.3847/1538-4357/aa72f1}

\bibitem[{{Nevins} \& {Roberts}(2023)}]{Nevins&Roberts23}
{Nevins}, B., \& {Roberts}, L.~F. 2023, \mnras, \dodoi{10.1093/mnras/stad372}

\bibitem[{{Otsuki} {et~al.}(2000){Otsuki}, {Tagoshi}, {Kajino}, \& {Wanajo}}]{Otsuki+00}
{Otsuki}, K., {Tagoshi}, H., {Kajino}, T., \& {Wanajo}, S.-y. 2000, \apj, 533, 424, \dodoi{10.1086/308632}

\bibitem[{{Paczynski}(1983)}]{Paczynski1983}
{Paczynski}, B. 1983, \apj, 267, 315, \dodoi{10.1086/160870}

\bibitem[{{Patat} {et~al.}(1994){Patat}, {Barbon}, {Cappellaro}, \& {Turatto}}]{Patat+94}
{Patat}, F., {Barbon}, R., {Cappellaro}, E., \& {Turatto}, M. 1994, \aap, 282, 731

\bibitem[{{Paxton} {et~al.}(2011){Paxton}, {Bildsten}, {Dotter}, {Herwig}, {Lesaffre}, \& {Timmes}}]{Paxton+11}
{Paxton}, B., {Bildsten}, L., {Dotter}, A., {et~al.} 2011, \apjs, 192, 3, \dodoi{10.1088/0067-0049/192/1/3}

\bibitem[{{Paxton} {et~al.}(2013){Paxton}, {Cantiello}, {Arras}, {Bildsten}, {Brown}, {Dotter}, {Mankovich}, {Montgomery}, {Stello}, {Timmes}, \& {Townsend}}]{Paxton+13}
{Paxton}, B., {Cantiello}, M., {Arras}, P., {et~al.} 2013, \apjs, 208, 4, \dodoi{10.1088/0067-0049/208/1/4}

\bibitem[{{Paxton} {et~al.}(2015){Paxton}, {Marchant}, {Schwab}, {Bauer}, {Bildsten}, {Cantiello}, {Dessart}, {Farmer}, {Hu}, {Langer}, {Townsend}, {Townsley}, \& {Timmes}}]{Paxton+15}
{Paxton}, B., {Marchant}, P., {Schwab}, J., {et~al.} 2015, \apjs, 220, 15, \dodoi{10.1088/0067-0049/220/1/15}

\bibitem[{{Paxton} {et~al.}(2018){Paxton}, {Schwab}, {Bauer}, {Bildsten}, {Blinnikov}, {Duffell}, {Farmer}, {Goldberg}, {Marchant}, {Sorokina}, {Thoul}, {Townsend}, \& {Timmes}}]{Paxton+18}
{Paxton}, B., {Schwab}, J., {Bauer}, E.~B., {et~al.} 2018, \apjs, 234, 34, \dodoi{10.3847/1538-4365/aaa5a8}

\bibitem[{Pian {et~al.}(2017)Pian, D'Avanzo, Benetti, Branchesi, Brocato, Campana, Cappellaro, Covino, D'Elia, Fynbo, Getman, Ghirlanda, Ghisellini, Grado, Greco, Hjorth, Kouveliotou, Levan, Limatola, Malesani, Mazzali, Melandri, M{\o}ller, Nicastro, Palazzi, Piranomonte, Rossi, Salafia, Selsing, Stratta, Tanaka, Tanvir, Tomasella, Watson, Yang, Amati, Antonelli, Ascenzi, Bernardini, Bo{\"e}r, Bufano, Bulgarelli, Capaccioli, Casella, Castro-Tirado, Chassande-Mottin, Ciolfi, Copperwheat, Dadina, De~Cesare, Di~Paola, Fan, Gendre, Giuffrida, Giunta, Hunt, Israel, Jin, Kasliwal, Klose, Lisi, Longo, Maiorano, Mapelli, Masetti, Nava, Patricelli, Perley, Pescalli, Piran, Possenti, Pulone, Razzano, Salvaterra, Schipani, Spera, Stamerra, Stella, Tagliaferri, Testa, Troja, Turatto, Vergani, \& Vergani}]{Pian+2017}
Pian, E., D'Avanzo, P., Benetti, S., {et~al.} 2017, Nature, 551, 67, \dodoi{10.1038/nature24298}

\bibitem[{{Piran} {et~al.}(2017){Piran}, {Nakar}, {Mazzali}, \& {Pian}}]{Piran+17}
{Piran}, T., {Nakar}, E., {Mazzali}, P., \& {Pian}, E. 2017, arXiv e-prints, arXiv:1704.08298, \dodoi{10.48550/arXiv.1704.08298}

\bibitem[{{Piro} {et~al.}(2021){Piro}, {Haynie}, \& {Yao}}]{Piro+21}
{Piro}, A.~L., {Haynie}, A., \& {Yao}, Y. 2021, \apj, 909, 209, \dodoi{10.3847/1538-4357/abe2b1}

\bibitem[{{Podsiadlowski} {et~al.}(1992){Podsiadlowski}, {Joss}, \& {Hsu}}]{Podsiadlowski+1992}
{Podsiadlowski}, P., {Joss}, P.~C., \& {Hsu}, J.~J.~L. 1992, \apj, 391, 246, \dodoi{10.1086/171341}

\bibitem[{{Popov}(1993)}]{Popov93}
{Popov}, D.~V. 1993, \apj, 414, 712, \dodoi{10.1086/173117}

\bibitem[{{Prasanna} {et~al.}(2022){Prasanna}, {Coleman}, {Raives}, \& {Thompson}}]{Prasanna+22}
{Prasanna}, T., {Coleman}, M. S.~B., {Raives}, M.~J., \& {Thompson}, T.~A. 2022, \mnras, 517, 3008, \dodoi{10.1093/mnras/stac2651}

\bibitem[{{Prasanna} {et~al.}(2023){Prasanna}, {Coleman}, {Raives}, \& {Thompson}}]{Prasanna+23}
---. 2023, \mnras, 526, 3141, \dodoi{10.1093/mnras/stad2948}

\bibitem[{{Qian} \& {Woosley}(1996)}]{Qian&Woosley96}
{Qian}, Y., \& {Woosley}, S.~E. 1996, \apj, 471, 331, \dodoi{10.1086/177973}

\bibitem[{Quataert {et~al.}(2019)Quataert, Lecoanet, \& Coughlin}]{Quataert+Coughlin2019}
Quataert, E., Lecoanet, D., \& Coughlin, E.~R. 2019, Monthly Notices of the Royal Astronomical Society: Letters, 485, L83, \dodoi{10.1093/mnrasl/slz031}

\bibitem[{{Rastinejad} {et~al.}(2022){Rastinejad}, {Gompertz}, {Levan}, {Fong}, {Nicholl}, {Lamb}, {Malesani}, {Nugent}, {Oates}, {Tanvir}, {de Ugarte Postigo}, {Kilpatrick}, {Moore}, {Metzger}, {Ravasio}, {Rossi}, {Schroeder}, {Jencson}, {Sand}, {Smith}, {Ag{\"u}{\'\i} Fern{\'a}ndez}, {Berger}, {Blanchard}, {Chornock}, {Cobb}, {De Pasquale}, {Fynbo}, {Izzo}, {Kann}, {Laskar}, {Marini}, {Paterson}, {Rouco Escorial}, {Sears}, \& {Th{\"o}ne}}]{Rastinejad+22}
{Rastinejad}, J.~C., {Gompertz}, B.~P., {Levan}, A.~J., {et~al.} 2022, arXiv e-prints, arXiv:2204.10864.
\newblock \doarXiv{2204.10864}

\bibitem[{{Rastinejad} {et~al.}(2023){Rastinejad}, {Fong}, {Levan}, {Tanvir}, {Kilpatrick}, {Fruchter}, {Anand}, {Bhirombhakdi}, {Covino}, {Fynbo}, {Halevi}, {Hartmann}, {Heintz}, {Izzo}, {Jakobsson}, {Lamb}, {Malesani}, {Melandri}, {Metzger}, {Milvang-Jensen}, {Pian}, {Pugliese}, {Rossi}, {Siegel}, {Singh}, \& {Stratta}}]{Rastinejad+23}
{Rastinejad}, J.~C., {Fong}, W., {Levan}, A.~J., {et~al.} 2023, arXiv e-prints, arXiv:2312.04630, \dodoi{10.48550/arXiv.2312.04630}

\bibitem[{{Renzo}(2015)}]{renzo:15}
{Renzo}, M. 2015, Master's thesis, Universit\`a di Pisa, Italy

\bibitem[{{Renzo} {et~al.}(2019){Renzo}, {Zapartas}, {de Mink}, {G{\"o}tberg}, {Justham}, {Farmer}, {Izzard}, {Toonen}, \& {Sana}}]{Renzo+19b}
{Renzo}, M., {Zapartas}, E., {de Mink}, S.~E., {et~al.} 2019, \aap, 624, A66, \dodoi{10.1051/0004-6361/201833297}

\bibitem[{{Rozwadowska} {et~al.}(2021){Rozwadowska}, {Vissani}, \& {Cappellaro}}]{Rozwadowska+21}
{Rozwadowska}, K., {Vissani}, F., \& {Cappellaro}, E. 2021, \na, 83, 101498, \dodoi{10.1016/j.newast.2020.101498}

\bibitem[{Sandoval {et~al.}(2021)Sandoval, Hix, Messer, Lentz, \& Harris}]{Sandoval+21}
Sandoval, M.~A., Hix, W.~R., Messer, O. E.~B., Lentz, E.~J., \& Harris, J.~A. 2021, The Astrophysical Journal, 921, 113, \dodoi{10.3847/1538-4357/ac1d49}

\bibitem[{{Siegel} {et~al.}(2021){Siegel}, {Agarwal}, {Barnes}, {Metzger}, {Renzo}, \& {Villar}}]{Siegel.ea_2021.arXiv_super.kilonovae}
{Siegel}, D.~M., {Agarwal}, A., {Barnes}, J., {et~al.} 2021, arXiv e-prints, arXiv:2111.03094.
\newblock \doarXiv{2111.03094}

\bibitem[{{Siegel} {et~al.}(2022){Siegel}, {Agarwal}, {Barnes}, {Metzger}, {Renzo}, \& {Villar}}]{Siegel+22}
---. 2022, \apj, 941, 100, \dodoi{10.3847/1538-4357/ac8d04}

\bibitem[{{Siegel} {et~al.}(2019){Siegel}, {Barnes}, \& {Metzger}}]{Siegel+19}
{Siegel}, D.~M., {Barnes}, J., \& {Metzger}, B.~D. 2019, \nat, 569, 241, \dodoi{10.1038/s41586-019-1136-0}

\bibitem[{{Simon} {et~al.}(2023){Simon}, {Brown}, {Mutlu-Pakdil}, {Ji}, {et~al.}}]{Simon+23}
{Simon}, J.~D., {Brown}, T.~M., {Mutlu-Pakdil}, B., {Ji}, A.~P., {et~al.} 2023, \apj, 944, 43, \dodoi{10.3847/1538-4357/aca9d1}

\bibitem[{{Smartt}(2009)}]{Smartt2009}
{Smartt}, S.~J. 2009, \araa, 47, 63, \dodoi{10.1146/annurev-astro-082708-101737}

\bibitem[{{Smith} {et~al.}(2011){Smith}, {Li}, {Filippenko}, \& {Chornock}}]{Smith+2011}
{Smith}, N., {Li}, W., {Filippenko}, A.~V., \& {Chornock}, R. 2011, \mnras, 412, 1522, \dodoi{10.1111/j.1365-2966.2011.17229.x}

\bibitem[{{Spergel} {et~al.}(2015){Spergel}, {Gehrels}, {Baltay}, {Bennett}, {Breckinridge}, {Donahue}, {Dressler}, {Gaudi}, {Greene}, {Guyon}, {Hirata}, {Kalirai}, {Kasdin}, {Macintosh}, {Moos}, {Perlmutter}, {Postman}, {Rauscher}, {Rhodes}, {Wang}, {Weinberg}, {Benford}, {Hudson}, {Jeong}, {Mellier}, {Traub}, {Yamada}, {Capak}, {Colbert}, {Masters}, {Penny}, {Savransky}, {Stern}, {Zimmerman}, {Barry}, {Bartusek}, {Carpenter}, {Cheng}, {Content}, {Dekens}, {Demers}, {Grady}, {Jackson}, {Kuan}, {Kruk}, {Melton}, {Nemati}, {Parvin}, {Poberezhskiy}, {Peddie}, {Ruffa}, {Wallace}, {Whipple}, {Wollack}, \& {Zhao}}]{Spergel+15}
{Spergel}, D., {Gehrels}, N., {Baltay}, C., {et~al.} 2015, arXiv e-prints, arXiv:1503.03757.
\newblock \doarXiv{1503.03757}

\bibitem[{{Spruit}(1999)}]{Spruit1999}
{Spruit}, H.~C. 1999, \aap, 349, 189, \dodoi{10.48550/arXiv.astro-ph/9907138}

\bibitem[{{Spruit}(2002)}]{spruit:02}
---. 2002, \aap, 381, 923, \dodoi{10.1051/0004-6361:20011465}

\bibitem[{{Spyromilio} {et~al.}(1990){Spyromilio}, {Meikle}, \& {Allen}}]{Spyr+1990}
{Spyromilio}, J., {Meikle}, W.~P.~S., \& {Allen}, D.~A. 1990, \mnras, 242, 669, \dodoi{10.1093/mnras/242.4.669}

\bibitem[{{Stello} {et~al.}(2016){Stello}, {Cantiello}, {Fuller}, {Huber}, {Garc{\'\i}a}, {Bedding}, {Bildsten}, \& {Silva Aguirre}}]{Stello+16}
{Stello}, D., {Cantiello}, M., {Fuller}, J., {et~al.} 2016, \nat, 529, 364, \dodoi{10.1038/nature16171}

\bibitem[{{Strotjohann} {et~al.}(2023){Strotjohann}, {Ofek}, {Gal-Yam}, {Sollerman}, {Chen}, {Yaron}, {Zackay}, {Rehemtulla}, {Gris}, {Masci}, {Rusholme}, \& {Purdum}}]{Strotjohann+23}
{Strotjohann}, N.~L., {Ofek}, E.~O., {Gal-Yam}, A., {et~al.} 2023, arXiv e-prints, arXiv:2303.00010, \dodoi{10.48550/arXiv.2303.00010}

\bibitem[{{Sukhbold} {et~al.}(2016){Sukhbold}, {Ertl}, {Woosley}, {Brown}, \& {Janka}}]{Sukhbold+16}
{Sukhbold}, T., {Ertl}, T., {Woosley}, S.~E., {Brown}, J.~M., \& {Janka}, H.~T. 2016, \apj, 821, 38, \dodoi{10.3847/0004-637X/821/1/38}

\bibitem[{{Sukhbold} \& {Thompson}(2017)}]{Sukhbold&Thompson17}
{Sukhbold}, T., \& {Thompson}, T.~A. 2017, \mnras, 472, 224, \dodoi{10.1093/mnras/stx2004}

\bibitem[{{Tanaka} \& {Hotokezaka}(2013)}]{Tanaka&Hotokezaka13}
{Tanaka}, M., \& {Hotokezaka}, K. 2013, Astrophys. J., 775, 113, \dodoi{10.1088/0004-637X/775/2/113}

\bibitem[{{Tanaka} {et~al.}(2020){Tanaka}, {Kato}, {Gaigalas}, \& {Kawaguchi}}]{Tanaka+20}
{Tanaka}, M., {Kato}, D., {Gaigalas}, G., \& {Kawaguchi}, K. 2020, \mnras, 496, 1369, \dodoi{10.1093/mnras/staa1576}

\bibitem[{Tanaka {et~al.}(2017)Tanaka, Utsumi, Mazzali, Tominaga, Yoshida, Sekiguchi, Morokuma, Motohara, Ohta, Kawabata, Abe, Aoki, Asakura, Baar, Barway, Bond, Doi, Fujiyoshi, Furusawa, Honda, Itoh, Kawabata, Kawai, Kim, Lee, Miyazaki, Morihana, Nagashima, Nagayama, Nakaoka, Nakata, Ohsawa, Ohshima, Okita, Saito, Sumi, Tajitsu, Takahashi, Takayama, Tamura, Tanaka, Terai, Tristram, Yasuda, \& Zenko}]{Tanaka+2017}
Tanaka, M., Utsumi, Y., Mazzali, P.~A., {et~al.} 2017, Publications of the Astronomical Society of Japan, 69, 102, \dodoi{10.1093/pasj/psx121}

\bibitem[{{Thielemann} {et~al.}(2020){Thielemann}, {Wehmeyer}, \& {Wu}}]{Thielemann+20}
{Thielemann}, F.-K., {Wehmeyer}, B., \& {Wu}, M.-R. 2020, in Journal of Physics Conference Series, Vol. 1668, Journal of Physics Conference Series, 012044, \dodoi{10.1088/1742-6596/1668/1/012044}

\bibitem[{{Thompson}(2003)}]{Thompson03}
{Thompson}, T.~A. 2003, \apjl, 585, L33, \dodoi{10.1086/374261}

\bibitem[{{Thompson} {et~al.}(2001){Thompson}, {Burrows}, \& {Meyer}}]{Thompson+01}
{Thompson}, T.~A., {Burrows}, A., \& {Meyer}, B.~S. 2001, \apj, 562, 887, \dodoi{10.1086/323861}

\bibitem[{{Thompson} {et~al.}(2004){Thompson}, {Chang}, \& {Quataert}}]{Thompson+04}
{Thompson}, T.~A., {Chang}, P., \& {Quataert}, E. 2004, \apj, 611, 380, \dodoi{10.1086/421969}

\bibitem[{{Thompson} \& {ud-Doula}(2018)}]{Thompson&udDoula18}
{Thompson}, T.~A., \& {ud-Doula}, A. 2018, \mnras, 476, 5502, \dodoi{10.1093/mnras/sty480}

\bibitem[{{Tonry} {et~al.}(2018){Tonry}, {Denneau}, {Heinze}, {Stalder}, {Smith}, {Smartt}, {Stubbs}, {Weiland}, \& {Rest}}]{Tonry+18}
{Tonry}, J.~L., {Denneau}, L., {Heinze}, A.~N., {et~al.} 2018, \pasp, 130, 064505, \dodoi{10.1088/1538-3873/aabadf}

\bibitem[{{Utrobin}(2007)}]{Utrobin2007}
{Utrobin}, V.~P. 2007, \aap, 461, 233, \dodoi{10.1051/0004-6361:20066078}

\bibitem[{{Utrobin} {et~al.}(2017){Utrobin}, {Wongwathanarat}, {Janka}, \& {M{\"u}ller}}]{Utrobin2017}
{Utrobin}, V.~P., {Wongwathanarat}, A., {Janka}, H.~T., \& {M{\"u}ller}, E. 2017, \apj, 846, 37, \dodoi{10.3847/1538-4357/aa8594}

\bibitem[{Valenti {et~al.}(2016)Valenti, Howell, Stritzinger, Graham, Hosseinzadeh, Arcavi, Bildsten, Jerkstrand, McCully, Pastorello, Piro, Sand, Smartt, Terreran, Baltay, Benetti, Brown, Filippenko, Fraser, Rabinowitz, Sullivan, \& Yuan}]{Valenti+2016}
Valenti, S., Howell, D.~A., Stritzinger, M.~D., {et~al.} 2016, Monthly Notices of the Royal Astronomical Society, 459, 3939, \dodoi{10.1093/mnras/stw870}

\bibitem[{{Villar} {et~al.}(2021){Villar}, {Cranmer}, {Berger}, {Contardo}, {Ho}, {Hosseinzadeh}, \& {Lin}}]{Villar+21}
{Villar}, V.~A., {Cranmer}, M., {Berger}, E., {et~al.} 2021, \apjs, 255, 24, \dodoi{10.3847/1538-4365/ac0893}

\bibitem[{{Vink} {et~al.}(2001){Vink}, {de Koter}, \& {Lamers}}]{Vink+2001}
{Vink}, J.~S., {de Koter}, A., \& {Lamers}, H.~J.~G.~L.~M. 2001, \aap, 369, 574, \dodoi{10.1051/0004-6361:20010127}

\bibitem[{{Vlasov} {et~al.}(2017){Vlasov}, {Metzger}, {Lippuner}, {Roberts}, \& {Thompson}}]{Vlasov+17}
{Vlasov}, A.~D., {Metzger}, B.~D., {Lippuner}, J., {Roberts}, L.~F., \& {Thompson}, T.~A. 2017, \mnras, 468, 1522, \dodoi{10.1093/mnras/stx478}

\bibitem[{{Wallner} {et~al.}(2015){Wallner}, {Faestermann}, {Feige}, {Feldstein}, {Knie}, {Korschinek}, {Kutschera}, {Ofan}, {Paul}, {Quinto}, {Rugel}, \& {Steier}}]{Wallner+15}
{Wallner}, A., {Faestermann}, T., {Feige}, J., {et~al.} 2015, Nature Commun., 6, 5956, \dodoi{10.1038/ncomms6956}

\bibitem[{{Wanajo}(2018)}]{Wanajo18}
{Wanajo}, S. 2018, \apj, 868, 65, \dodoi{10.3847/1538-4357/aae0f2}

\bibitem[{{Wang} \& {Burrows}(2023)}]{Wang&Burrows23}
{Wang}, T., \& {Burrows}, A. 2023, arXiv e-prints, arXiv:2306.13712, \dodoi{10.48550/arXiv.2306.13712}

\bibitem[{{Wheeler} \& {Chatzopoulos}(2023)}]{Wheeler&Chatzaopoulos23}
{Wheeler}, J.~C., \& {Chatzopoulos}, E. 2023, Astronomy and Geophysics, 64, 3.11, \dodoi{10.1093/astrogeo/atad020}

\bibitem[{{Winteler} {et~al.}(2012){Winteler}, {K{\"a}ppeli}, {Perego}, {Arcones}, {Vasset}, {Nishimura}, {Liebend{\"o}rfer}, \& {Thielemann}}]{Winteler+12}
{Winteler}, C., {K{\"a}ppeli}, R., {Perego}, A., {et~al.} 2012, Astrophys. J. Lett., 750, L22, \dodoi{10.1088/2041-8205/750/1/L22}

\bibitem[{{Wongwathanarat} {et~al.}(2015){Wongwathanarat}, {M{\"u}ller}, \& {Janka}}]{Wongwathanarat+15}
{Wongwathanarat}, A., {M{\"u}ller}, E., \& {Janka}, H.~T. 2015, \aap, 577, A48, \dodoi{10.1051/0004-6361/201425025}

\bibitem[{{Woosley} {et~al.}(1994){Woosley}, {Wilson}, {Mathews}, {Hoffman}, \& {Meyer}}]{Woosley+94}
{Woosley}, S.~E., {Wilson}, J.~R., {Mathews}, G.~J., {Hoffman}, R.~D., \& {Meyer}, B.~S. 1994, \apj, 433, 229, \dodoi{10.1086/174638}

\bibitem[{Zapartas {et~al.}(2017)}]{Zapartas+17a}
Zapartas, E., {et~al.} 2017, Astron. Astrophys., 601, A29, \dodoi{10.1051/0004-6361/201629685}

\end{thebibliography}

\appendix

\section{Figure 5 Adapted for \textit{LSST}} \label{app:LSSTfig}
\begin{figure*}[h]
    \centering
    \includegraphics[width=1.0\textwidth]{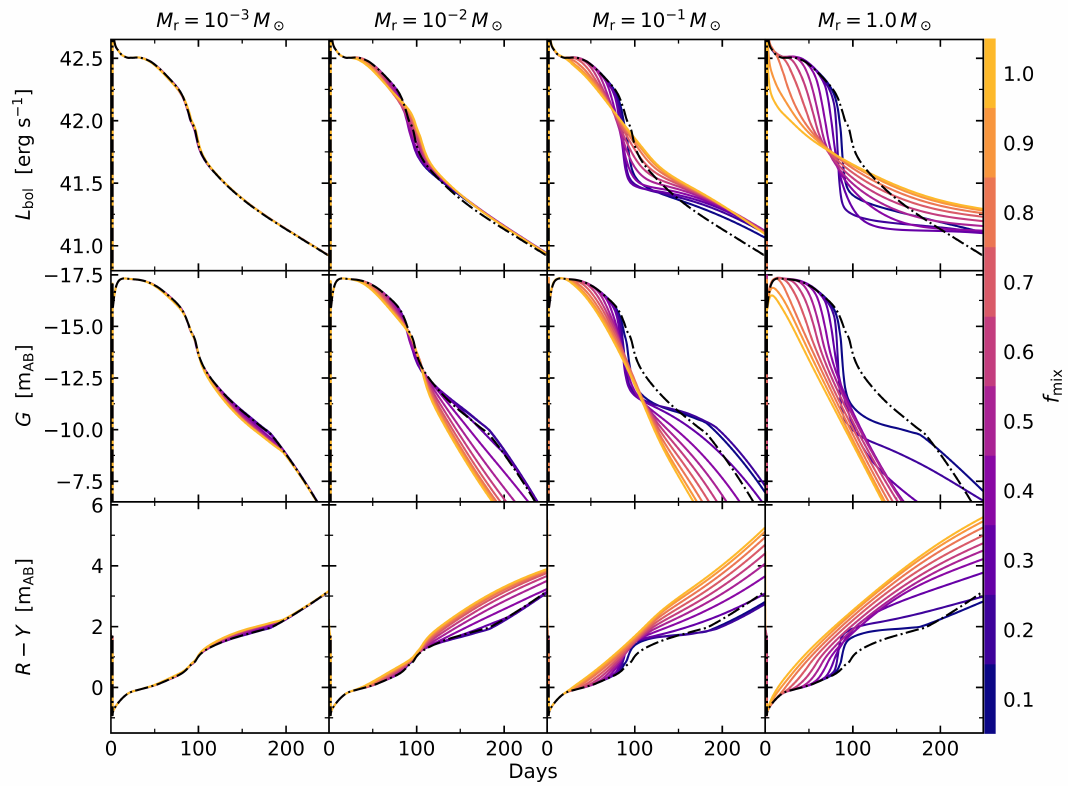}
    \caption{\normalsize Same as Fig.~\ref{fig:grid}, except with band filters substituted to those used in LSST. $V$-band is changed to $G$ and $R-K$ index is changed to $R-Y$.}
    \label{fig:gridLSST}
\end{figure*}

To better facilitate comparison with observations, Fig.~\ref{fig:gridLSST} shows a revised version of Fig.~\ref{fig:grid} with photometric bands replaced to those that are available to \textit{LSST}.

\section{Model Tests}
\label{sec:tests}

Here we describe several numerical tests of our models, to assess the robustness of our conclusions in face of different assumptions regarding the opacity floor and the role of energy input from $r$-process radioactive decay.

\subsection{Default Opacity Floor}
\label{app:defaultfloor}

According to \citet{Morozova15}, the ``default'' opacity floor profile in \texttt{SNEC}, set such that $\kappa_{\rm f}(M_{\rm NS}) = 0.24$ cm$^2$ g$^{-1}$ and $\kappa_{\rm f}(M_{\rm ej}) = 0.01$ cm$^2$ g$^{-1}$, was chosen so as to produce light-curves consistent with previous modeling \citep{Blinnikov+Bartunov93, Blinnikov+98, Bersten2011}. For consistency, we have thus adopted this default opacity prescription for the non-$r$-process-enriched material in all models presented thus far in this paper. However, given that these values are not motivated by first principles, they may overestimate the true physical opacities of the system.

\begin{figure}
    \centering
    \includegraphics[width=0.8\textwidth]{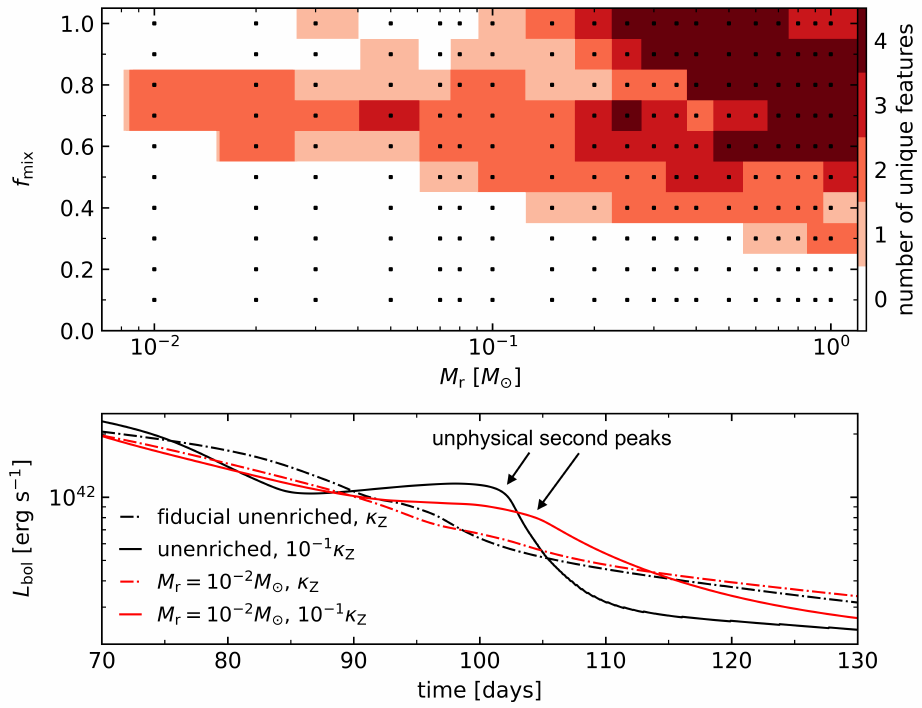}
    \caption{\normalsize Unenriched and $M_{\rm r} = 10^{-2} M_{\odot}$ enriched light-curves with a 'default' opacity floor profile reduced by a factor of 10 to $10^{-1} \kappa_{\rm Z}$. The corresponding fiducial floor, $\kappa_{\rm Z}$, light-curves are included for comparison. We see an unphysical double peaked behavior in the low floor unenriched case. The inclusion of $r$-process slightly reduces the onset of the second peak, however, the feature is still significant at low levels of enrichment.}
    \label{fig:floor_test}
\end{figure}

Since an unphysically high baseline opacity would ``dilute'' the opacity effects of $r$-process elements, we explore the sensitivity of our results to the floor opacity treatment. Repeating an otherwise identical analysis by reducing the floor opacity by a factor of 10 (such that $\kappa_{\rm f}(M_{\rm NS}) = 0.024$ cm$^2$ g$^{-1}$ and $\kappa_{\rm f}(M_{\rm ej}) = 0.001$ cm$^2$ g$^{-1}$) we find that the detectability threshold for $r$-process enrichment is extended down to lower levels, $M_{\rm r} \sim 10^{-2} M_{\odot}$ and $f_{\rm mix} \lesssim 0.6$, than in the high-floor case, as shown in the top panel of Fig.~\ref{fig:floor_test}.

However, as shown in the bottom panel of the same figure, the implementation of this low floor opacity also introduces strange, likely unphysical light-curve features around $\approx 100$ days. These include a second light-curve peak, which arises at least in part from those steep gradients in the radial density profile of the ejecta shown in Fig.~\ref{fig:initialfiducial} (see, e.g. discussions in \citealt{Eastman1994, Utrobin2007}, and in particular, \citealt{Utrobin2017}). Such density profile features may themselves be unphysical, as multidimensional effects and hydrodynamic instabilities during the early shock-crossing explosion phase, such as the Rayleigh-Taylor Instability \citep{Herant1994,Wongwathanarat+15, Duffell2016, Paxton+18} would smooth these out in more realistic 3D simulations.

Due to uncertainties in how the ``detectable" signatures of $r$-process enrichment may be confounded with these unphysical effects on the light-curve, we opt not to draw strong conclusions from this analysis with the low floor opacity. We maintain that the results with the fiducial floor provide a conservative estimate for the detectability boundary.

\subsection{Effects of $r$-Process Heating}
\label{app:heatingeffects}

\begin{figure}
    \centering
    \includegraphics[width=0.8\textwidth]{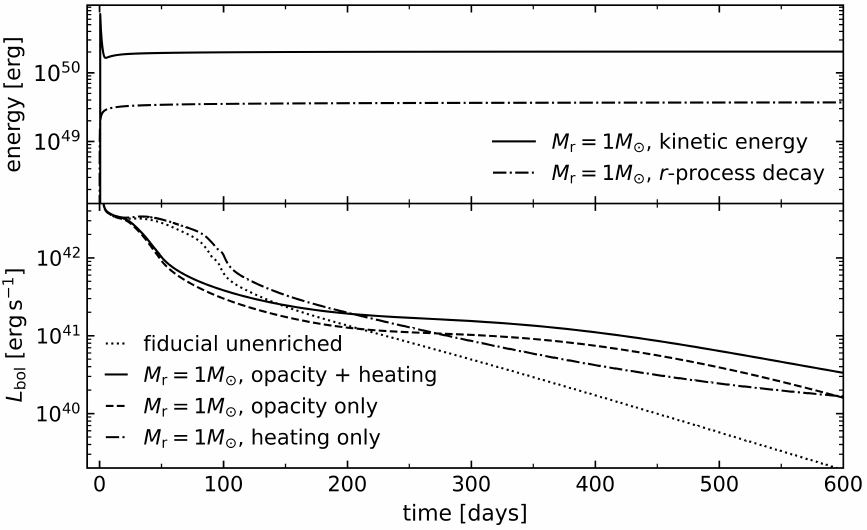}

    \caption{\normalsize The top panel shows the total kinetic energy of the $r$-process-enriched layers (from $M_{\rm NS} = 1.4 M_{\odot} \text{ to } M_{\rm r,mix} = 9.0 M_{\odot}$) and the total energy radiated from decay of $r$-nuclei in these layers for an extremely enriched ($M_{\rm r}=1.0M_\odot$, $f_{\rm mix}=0.7$) model. The bottom panel compares the light-curves of three models with the same quantity and distribution of $r$-process: $r$-heating and $r$-opacity, $r$-opacity only, and $r$-heating only. An otherwise equivalent unenriched model is shown for comparison.}
    \label{fig:heating}
\end{figure}

Although our models include heating due to the decay of $r$-process nuclei, we do not expect its presence to result in significant changes to the dynamical evolution of the ejecta nor the observed luminosity, for reasons outlined here. Regarding dynamics, the top panel of Fig.~\ref{fig:heating} shows as a function of time the integrated energy input from $r$-process decay, $E_{\rm r}(t)$, in comparison to the total ejecta kinetic energy, $E_{\rm k}(t)$, for a highly $r$-process-enriched model with $M_{\rm r} = 1 M_{\odot}$ (greater than the fiducial $M_{\rm r} = 0.3 M_\odot)$. These are calculated as,
\be
E_{\rm r}(t) = \int\limits_{1\,\mathrm{s}}^{t}\epsilon_{rp}(t')dt'\int\limits_{M_{\rm NS}}^{M_{\rm r, mix}}X_r(M)dM;\,\,\,
E_{\rm k}(t) = \int\limits_{M_{\rm NS}}^{M_{\rm r,mix}}\frac{1}{2}[v(M,t)]^2dM.
\ee
Even for this extreme level of enrichment, $E_{\rm r}$ is an order of magnitude smaller than $E_{\rm k}$, resulting in a negligible impact on the ejecta dynamics, even for complete conversion of the former into the latter via PdV work during the ejecta expansion.

 Regarding the effect of $r$-process decay on the SN luminosity budget, \citet{Siegel+19} showed that the specific energy deposition rate from the $^{56}$Ni decay chain exceeds that from $r$-process decay by $\sim 2$ orders of magnitude on the timescales of interest ($\lesssim 150$ days). For nickel masses $M_{\rm Ni} = 0.01 - 0.1 M_{\odot}$ relevant to SNe IIP, we see that $r$-process masses obeying $M_{\rm r} \lesssim 0.1 M_{\odot}$ should not appreciably alter the light-curve.

 For extreme enrichment events ($M_{\rm r} \sim 1 M_{\odot}$), the $r$-process decay luminosity can compete with $^{56}$Ni. This is evident in the ``heating only'' model in the bottom panel of Fig.~\ref{fig:heating}, whose light-curve deviates from the otherwise equivalent unenriched model. The ``heating only'' model accounts only for the effects of $r$-process heating. When the effects of higher $r$-process opacity are also included, as in the ``opacity + heating'' model, there is also an increase in the luminosity compared to the ``opacity only" model. The high opacity traps the energy from $r$-process decay, preventing it from escaping rapidly enough to appreciably alter the light-curve. This effect is most evident at late times, where the ``heating only" model is substantially more luminous than the unenriched model, while the differences between the ``opacity + heating" and the ``opacity only" models are less significant. Even for this extreme level of enrichment, $r$-process heating increases the luminosity by a factor of $\lesssim 2$ on timescales of interest, and the effect is further diminished for lower levels of enrichment. We conclude that $r$-process heating effects are minor, leaving opacity as the most prominent signature of $r$-process enrichment.

\subsection{Sensitivity to $\chi$}
\label{app:chi_effects}

\begin{figure}
    \centering
    \includegraphics[width=0.8\textwidth]{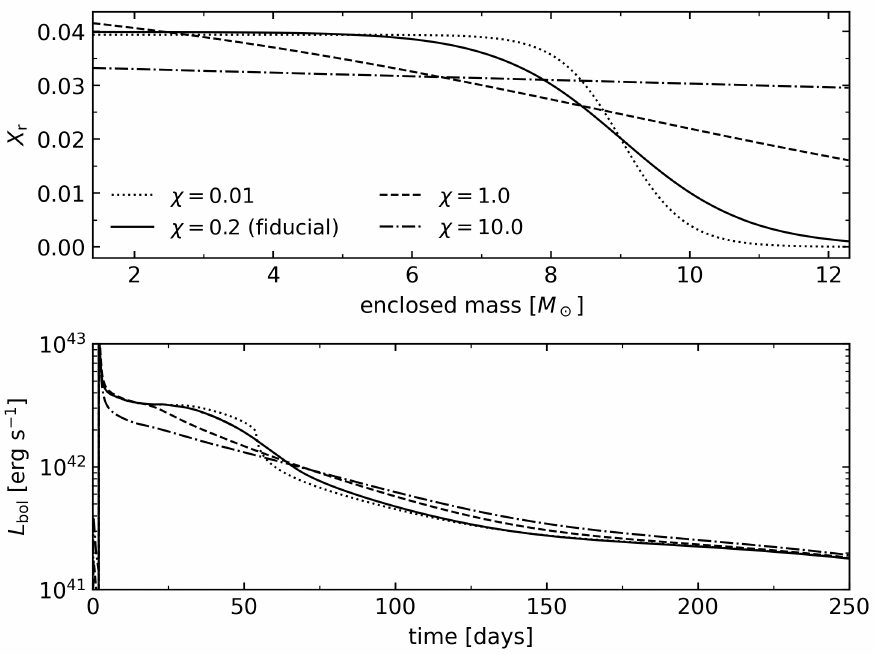}\label{fig:chi_test}
    \caption{\normalsize Mass fraction profiles (top panel) and light-curves (bottom panel) for models with the same $r$-process enrichment $\{M_{\rm r} = 0.3M_\odot,\ f_{\rm mix} = 0.7 \}$ but different values of $\chi$. The value $\chi = 0.2$ gives a sufficiently steep transition from $r$-process-enriched inner ejecta layers to the $r$-process free outer layers without introducing abrupt features in the light-curve (as in the $\chi = 0.01$ model at $\approx 60$ days).}
\end{figure}

The parameter $\chi$ entering Eq.~\ref{eq:Xr} was set to $\chi = 0.2$ in all $r$-process-enriched models presented thus far. The light-curves in Fig.~\ref{fig:chi_test} (bottom panel), corresponding to $r$-process mass fraction profiles of different $\chi$ values (top panel), show that our results are not overly sensitive to the value of $\chi$, as long as there is a steep transition from $r$-process-enriched layers to $r$-process free layers (e.g. $\chi = 0.01$, $\chi =0.2$).
Values of $\chi$ which are very high (e.g. $\chi=1,10$) entail $r$-rich material distributed far out into the ejecta, leading to degenerate behavior with a higher mixing fraction. Despite uncertainties in the exact distribution of $r$-process material, these degeneracies imply that results for varying $f_{\rm mix}$ can be translated to models with different choices for $\chi$. We take the smoother $\chi = 0.2$ as the fiducial value as it does not produce an abrupt change in the light-curve such as that seen in the $\chi = 0.01$ model at $\approx$ day 60.

\section{Light-Curve Morphology Fitting} \label{app: fitting}

\subsection{Fitting $t_{\rm p}$}

To estimate the plateau duration, $t_{\rm p}$, we fit the bolometric light-curve to the functional form:
\be
\mathrm{log}_{10}(L_{\rm bol}) = \frac{-A_0}{1+e^{(t-t_{\rm p})/W_0}} + P_0 t + M_0,
\label{tp_fit}
\ee
as introduced by \citet{Valenti+2016}. Following \citet{Goldberg+19}, we begin fitting the light-curves at the time  $ t = 0.75t_{\rm sd}$ where $t_{\rm sd}$ is the time of steepest descent in the bolometric light-curve (i.e., where the absolute value of the slope is maximum, excluding the initial rise to peak luminosity). The steepest descent typically occurs during the transition from the plateau to the $^{56}$Ni tail. We end the fit at $t = 1.75 t_{\rm sd}$. While we find this works well for unenriched models with sufficiently long plateaus $t_{\rm p} \approx t_{\rm sd} \approx 100$ days, the enriched models exhibit shorter plateaus, such that $t_{\rm sd} \lesssim 50$ days for highly enriched cases. Insofar that a start time $0.75 t_{\rm sd}$ is in these cases insufficient to encompass enough of the light-curve to produce a good fit, we add conditional arguments in our fitting procedure as follows: If $0.75 t_{\rm sd} < 10$ days, then $t_{\rm p} = 0$. If $0.75 t_{\rm sd} < 20$ days, the fit is started at day 10. If $0.75 t_{\rm sd} < 30$ days, the fit is started at day 15. If $0.75 t_{\rm sd} < 40$ days, the fit is started at day 20. If $0.75 t_{\rm sd} < 55$ days, the fit is started at day 30. If $t_{\rm sd} < 70$ days, the fit is started at 40 days. For $t_{\rm sd} > 70$ days, the start time is determined as in the standard case, at $0.75 t_{\rm sd}$. These conditions allow sufficient range in data to encompass all the necessary features of plateau, transition, and Ni-tail. We find good agreement between the results of this modified fitting algorithm and plateau durations estimated by eye, for models across all levels of enrichment.

\subsection{Fitting $s$}
In the observational literature, the plateau decline rate in $V$-band luminosity is typically divided into two components, $s_1$ and $s_2$, where $s_1$ is the slope magnitude of the initial steeper portion of the plateau (representing initial cooling from peak luminosity) and $s_2$ is the slope magnitude of the second, shallower section of the plateau as defined in \citealt{Anderson+14}. These quantities are defined to fit data from observations. The $V$-band light-curves produced by post-processing our \texttt{SNEC} bolometric luminosities do not show these two distinct slopes during the plateau. Even if the code could accurately simulate these features of the light-curve in unenriched models, it is uncertain how $r$-process enrichment would affect these features and whether two slopes would be distinguishable across all levels of enrichment. To accommodate these uncertainties while maintaining synonymity to the original definition of $s_2$, we consider for simplicity a single plateau decline rate $s$ by performing a linear fit starting at peak $V$-band magnitude and ending 35 days past peak, where $V$-band magnitude is calculated using the AB magnitude system assuming a black body at temperature $T_{\rm ph}$. The resulting decline rates accurately capture the trends visible by eye, where models with greater explosion energies or greater levels of $r$-process enrichment have steeper ``plateau" slopes.

\section{Modifications to \texttt{SNEC}}

As described in Sec.~\ref{sec:model}, we include the effects of $r$-process enrichment on the opacity of the SN ejecta by modifying the default opacity floor present in \texttt{SNEC}. To do so, we first determine the ejecta mass $M_{\rm ej}$ by subtracting the excised mass (corresponding to the central PNS) from the total pre-SN progenitor mass and then we calculate $M_{\rm r,mix}$ following Eq.~\eqref{fmix}, determining the normalization constant from the integral described in Sec.~\ref{sec:model}. For each grid point, \texttt{(i=1, imax=1000)}, the $r$-process mass-fraction $X_{\rm r}(\texttt{i})$ is evaluated as a function of $M(\texttt{i})$ as in Eq.~\eqref{eq:Xr} and the new $r$-process opacity floor, $\kappa_{\rm r}(\texttt{i})$, is calculated by Eq.~\eqref{eq:kappar}. The original (``default") opacity floor of each zone is reassigned to a new variable $\kappa_{\rm Z}(\texttt{i})$, while the native \texttt{SNEC} variable $\kappa_{\rm f}(\texttt{i})$ (which previously denoted the default floor), is now set as the sum of $\kappa_{\rm Z}(\texttt{i})$ and $\kappa_{\rm r}(\texttt{i})$.

\begin{figure}
    \centering
    \includegraphics[width=0.8\textwidth]{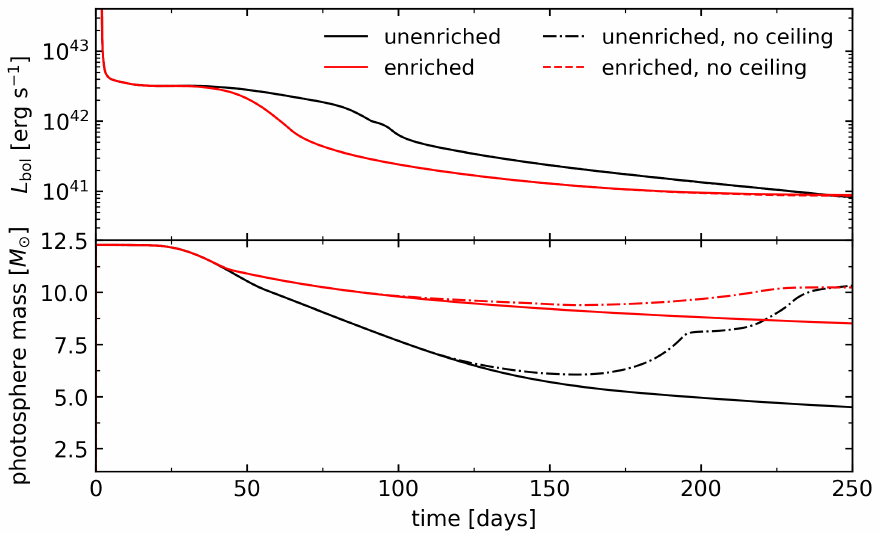}
    \caption{\normalsize Bolometric light-curves (top panel) and mass coordinate of the photosphere (bottom panel) for highly-enriched (red) and unenriched (black) SN models with (solid) and without (dashed) an opacity ceiling at low temperature. While the evolution of the bolometric light-curve is insensitive to our inclusion of an opacity ceiling at low temperatures, the photosphere evolution and hence the observed emission temperature is greatly affected.}
    \label{fig:floorceiling}
\end{figure}

\texttt{SNEC} refers to the OPAL \citep{Iglesias&Rogers1996} and Ferguson \citep{Ferguson2005} opacity tables to determine the ejecta opacity as a function of the temperature and density in a given grid cell. OPAL opacities  are used at high temperatures ($10^{3.75}$ K$ < T < 10^{8.7}$ K) while Ferguson opacities are used at low temperatures ($10^{2.7}$ K$ < T < 10^{4.5}$ K) \citep{Morozova15}. In the overlapping region, preference is given to Ferguson. If the opacity value read from the table for zone \texttt{i} is less than the corresponding opacity floor for that zone, then the opacity is set to the floor value instead of the table value. At low temperatures ($T < 2700$ K), the Ferguson opacities rise to high values due to the assumed presence of dust.  This leads to a sudden increase in the opacity of the outermost zones in the explosion at late time, which causes the photosphere to move outwards in mass coordinate at late times (``no ceiling" models, shown in Fig.~\ref{fig:floorceiling}). However, if dust formation is not efficient throughout the ejecta, then these opacities are over-estimated and such behavior is not physical (the photosphere radius should actually continue to move inwards monotonically). To circumvent this issue, at low temperatures ($T < 2700$ K) we artificially force opacities to the floor value, rather than using the opacity table; in effect, we use the opacity floor also as an opacity ``ceiling". We accomplish this by including a conditional argument that sets a zone's opacity value to the floor value if its temperature is less than 3000 K. To test the sensitivity of our early-time results to this assumed temperature threshold, we ran an otherwise identical model but with the ceiling instead set to trigger below 2000 K; reassuringly, the photosphere and luminosity evolution were found to be nearly identical to the 3000 K threshold case.

The variable \texttt{kappa} denotes the opacity value set by the opacity floor and ceiling, while \texttt{kappa\textunderscore table} denotes that assigned by the opacity tables. By default, the radial optical depth $\tau(M)$ (and hence the photosphere location) are calculated in \texttt{SNEC} using \texttt{kappa\textunderscore table}. In order to account for the opacity floor and ceiling limits on the photosphere profiles, we calculate $\tau$ using \texttt{kappa} instead of \texttt{kappa\textunderscore table}. With this modification and those described above, we are able to achieve the desired opacity profile (Fig.~\ref{fig:timelapse}) and avoid an unphysical outward expansion of the photosphere in mass coordinate. The corrected photosphere evolution for a $M_{\rm r} = 1 M_\odot$ enriched and fiducial unenriched models is shown in Fig.~\ref{fig:floorceiling}.

We include the effects of radioactive decay energy by multiplying the specific heating rate given by Eq.~\eqref{eq:epsilon_rp} with the $r$-process mass in each zone. We amend the definition of total luminosity to include this contribution, which thus becomes the sum of the photosphere luminosity, $^{56}$Ni decay luminosity external to the photosphere, and $r$-process decay luminosity external to the photosphere. We similarly amend the the ejecta heating rate internal to the photosphere.

\end{document}